\documentclass[sigconf]{acmart}
\acmConference[EASE 2025]{The 29th International Conference on Evaluation and Assessment in Software Engineering}{17–20 June, 2025}{Istanbul, Türkiye}

\acmMonth{06}
\acmYear{2025}
\copyrightyear{2025}

\usepackage{graphicx}
\usepackage[utf8]{inputenc}
\usepackage{algorithm}
\usepackage[noend]{algpseudocode}
\usepackage{multirow}
\usepackage{multicol}
\usepackage{enumitem}
\usepackage{ulem}
\usepackage{textcomp}
\usepackage{xcolor}
\usepackage{url}
\usepackage{balance}
\usepackage{array}
\usepackage{varwidth}

\title[FetaFix: Automatic Fault Localization and Repair of Deep Learning Model Conversions]{FetaFix: Automatic Fault Localization and Repair \\ of Deep Learning Model Conversions}

\author{Nikolaos Louloudakis}
\email{n.louloudakis@ed.ac.uk}
\affiliation{%
  \institution{University of Edinburgh}
  \country{United Kingdom}}
  
\author{Perry Gibson}
\email{perry.gibson@glasgow.ac.uk}
\affiliation{%
  \institution{University of Glasgow}
  \country{United Kingdom}}

\author{Jos\'e Cano}
\email{jose.canoreyes@glasgow.ac.uk}
\affiliation{%
  \institution{University of Glasgow}
  \country{United Kingdom}}

\author{Ajitha Rajan}
\email{arajan@ed.ac.uk}
\affiliation{%
  \institution{University of Edinburgh}
  \country{United Kingdom}}

\newcommand{\tool}{\texttt{FetaFix}} 
\newcommand{\source}{\textit{Source}}
\newcommand{\target}{\textit{Target}}

\begin{document}

\begin{abstract}

Converting deep learning models between frameworks is a common step to maximize model compatibility across devices and leverage optimization features that may be exclusively provided in one deep learning framework. 
However, this conversion process may be riddled with bugs, making the converted models either undeployable or problematic, considerably degrading their prediction correctness. 

In this paper, we propose an automated approach for fault localization and repair, \tool, during model conversion between deep learning frameworks. \tool\ is capable of detecting and fixing faults introduced in model input, parameters, hyperparameters, and the model graph during conversion.
\tool\ uses a set of fault types (mined from surveying common conversion issues reported in code repositories and forums) to localize potential conversion faults in the converted target model and then repair them appropriately, e.g., replacing the parameters of the target model with those from the source model. This is done iteratively for every image in the dataset, comparing output label differences between the source model and the converted target model until all differences are resolved.  
We evaluate the effectiveness of \tool\ in fixing model conversion bugs of three widely used image recognition models converted across four different deep learning frameworks. Overall, \tool\ was able to fix $462$ out of $755$ detected conversion faults, either completely repairing or significantly improving the performance of $14$ out of the $15$ erroneous conversion cases.

\end{abstract}

\maketitle

\vspace{-5pt}
\section{Introduction}
With the widespread use of deep learning (DL) in various domains, there is an inherent need for DL models to be inter-operable across DL frameworks (such as PyTorch~\cite{pytorch}, TensorFlow (TF)~\cite{tensorflow2015-whitepaper}, Keras~\cite{chollet2015keras}) to maximize re-usability of models across frameworks. Conversion of DL models between frameworks is facilitated by a plethora of automated conversion tools such as \texttt{tf2onnx}~\cite{tf2onnx}, \texttt{onnx2keras}~\cite{onnx2keras}, \texttt{onnx2torch}~\cite{onnx2torch},  \texttt{MMdnn}~\cite{MMdnn}, among others. 
However, the conversion process may be riddled with bugs~\cite{louloudakis2023deltann, louloudakis2022assessing,TF-errors}, making the converted models undeployable, perform poorly in terms of output label correctness when changing from one framework to another, run slowly, or face challenges in robust deployment~\cite{openja2022empirical, chen2020comprehensive,chen2021empirical}.

\begin{figure}[t]
     \centering
     \includegraphics[width=\columnwidth]{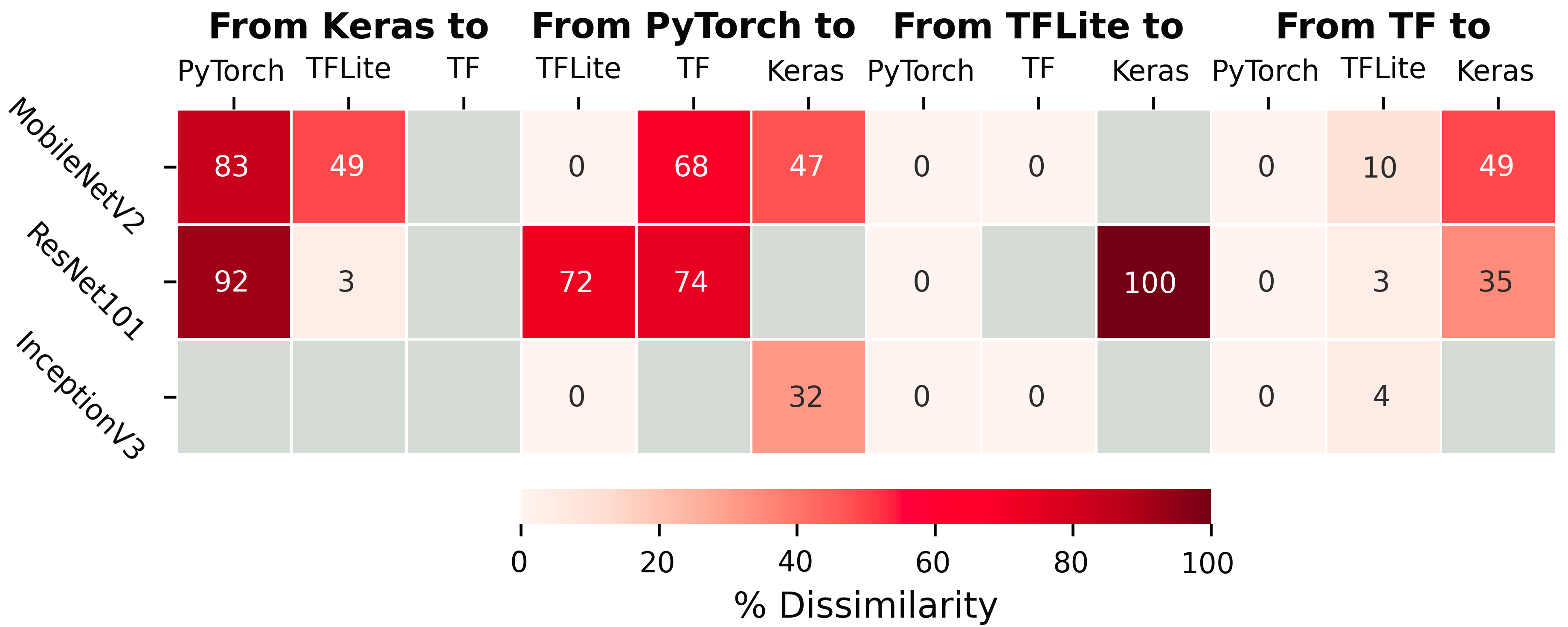}
     \vspace{-20pt}
     \caption{Pairwise comparison of output labels between \textit{Source} and converted \textit{Target} models (from~\cite{louloudakis2023deltann}).}
     \label{fig:conversionsorig}
     \vspace{-15pt}
\end{figure}

In a previous study of ours~\cite{louloudakis2023deltann}, we showcased empirical evidence of problems in model conversion across four well-known DL frameworks. In particular, we considered three image recognition models, pre-trained on ImageNet~\cite{deng2009imagenet}, that were treated as \textit{Source} models. We then converted each \textit{Source} model to use a different DL framework, referred to as \textit{Target}. 
Figure~\ref{fig:conversionsorig} from~\cite{louloudakis2023deltann} shows the proportion of output label dissimilarities between \textit{(Source, Target)} pairs across all images in the test dataset. 
The conversion process failed to execute in $11$ out of the $36$ conversions, seen as empty gray cells in Figure~\ref{fig:conversionsorig}. For the remaining conversions, they found varying levels of image label discrepancies between \textit{Target} and \textit{Source} models from 0 to 100\%, with $15$ cases having non-zero percentage of output label dissimilarity. This study underscores the error-prone nature of the conversion process, highlighting the necessity for a technique to localize and repair faults introduced by DL framework converters.

To address this problem, in this paper we propose \tool\footnote{The source code (accompanied with detailed instructions),  as well a small dataset for evaluation are available at \emph{\url{https://github.com/luludak/FetaFix}}}, an automated approach for fault localization and repair of erroneous model conversions between DL frameworks.
We focus on the cases in Figure~\ref{fig:conversionsorig} that did not crash, but resulted in non-zero output label differences across \source\ and \target. 
The gray boxes in Figure~\ref{fig:conversionsorig}, where the conversion tools crashed, have been explored by existing tools like MMdnn~\cite{MMdnn} and complements the capability we provide in diagnosing and repairing conversions with output label differences.

\tool\ is capable of detecting and fixing faults introduced in model layer weights, biases, hyperparameters, and the model computation graph during conversion. It compares the \source\ model (in one DL framework) against the converted \target\ model (in a different framework) using an input dataset. 
For images with label discrepancies between \source\ and \target, \tool\ assesses the difference between the models with respect to their parameters.
Then it uses a set of strategies to mitigate conversion errors such as the replacement of \target\ model parameters with those from \source.

Our approach for \tool\ focuses on the \target\ model rather than the converter tool itself for a variety of reasons. 
First, errors observed in conversions may be related to problematic configurations (e.g., input preprocessing) that are not related to the conversion tool.
Second, the source code of the conversion tool may be unavailable or proprietary, and therefore not accessible for direct repair.
Third, the conversion process might involve manual intervention or conversion (e.g., \cite{manualconversion1, manualconversion2}), which in turn can introduce errors.
Fourth, our methodology is not tied to any specific conversion tool/version, and can be applied to any model conversion between source and target settings, including intermediate representations like ONNX~\cite{onnxsite}. It is worth noting that conversion tools are constantly evolving, thus having a localize/repair approach independent of the tool source code accessibility and evolution would prove helpful and more dependable for AI model users.
As an aside, the fault localization and repair techniques we propose can also be adopted by conversion tool developers to address faults associated with the tools themselves.

Focusing on the non-crashing conversions from our prior work~\cite{louloudakis2023deltann}, we evaluated the effectiveness of \tool\ in fixing model conversion errors. \tool\ was able to effectively localize and repair $14$ of the $15$ erroneous model conversions that had non-zero output label dissimilarities.  

\noindent In summary, we make the following contributions:

\begin{enumerate}[topsep=0pt, itemsep=0pt, partopsep=0pt]
    \item A novel method for localizing faults in DL model conversion between a range of DL frameworks.
    
    \item An automated fault repair framework, \tool, that can fix converted \target\ models to match output labels with that of the \source\ model.
    
    \item A comprehensive empirical evaluation of the effectiveness of \tool\ in fault localization and repair. 
\end{enumerate}

\vspace{-5pt}
\section{Background}
We provide a brief background on deep learning frameworks, conversions between them, the intermediate ONNX format and the compiler framework Apache TVM that we use for implementation and analysis with \tool.

\vspace{-8pt}
\subsection{Deep Learning Frameworks}

Deep Learning Frameworks provide utilities to machine learning engineers and developers in order to build, train, deploy and optimize Deep Neural Networks (DNNs). We focus on four actively maintained and widely-used DL frameworks in our experiments: \textit{Keras}~\cite{chollet2015keras}, \textit{PyTorch}~\cite{pytorch}, \textit{TensorFlow (TF)}~\cite{tensorflow2015-whitepaper}, and \textit{TensorFlow Lite (TFLite} - a framework focused on edge devices)~\cite{tensorflow2015-whitepaper}.


\vspace{-8pt}
\subsection{Framework Conversions}

DNN models are converted from one DL framework to another, for two primary reasons: (1) portability - so that a model can be deployed on devices of different computational capabilities; and (2) supported features and/or operations across the Deep Learning Acceleration Stack~\cite{gibson_dlas_2025}, such as extended optimization support or better profiling options.
The conversion process is conducted by providing a pre-trained model built on one DL framework (e.g., PyTorch). We call this model \source. 
Then, the conversion tool will use \source\ in order to generate the same model, but built in a new DL framework (e.g., Keras). We call this model \target. 
There is a wide variety of open-source tools for this purpose~\cite{tf2onnx, onnx2keras, onnx2torch, MMdnn}, and the conversion procedure is performed automatically. 
However, the conversion process might involve multiple converters and might also require the model to be built using an intermediate representation (e.g., ONNX~\cite{onnxsite}). For instance, the conversion of a model from PyTorch to Keras might require a two-step conversion: (1) PyTorch to ONNX, using the native converter of PyTorch to generate ONNX; and (2), the \texttt{onnx2keras}~\cite{onnx2keras} converter, in order to convert the ONNX intermediate representation to Keras.

\vspace{-8pt}
\subsection{Open Neural Network Exchange (ONNX)}

ONNX~\cite{onnxsite} is an open standard for machine learning interoperability that allows the representation of DNNs using a common set of operators and file formats to enable interoperability with a variety of tools, and systems. 
ONNX provides a powerful API in order to analyze model graph, hyperparameters, and parameters. The model is represented by a graph of operations (or nodes), each containing a set of properties. These operations are executed sequentially, propagating data from one layer to its subsequent ones. Throughout the years, ONNX evolves producing newer versions of the standard, modified from the past. Each standard version is attributed an \texttt{opset} number.
Due to the popularity of ONNX and the versatility of its API, we perform our automatic fault localization and repair in this representation level - retrieving ONNX versions of the \source\ and \target\ model, analyzing them and repairing them. For that purpose, when we refer to \source\ and \target\, we implicitly refer to the ONNX variants of the models under test.

\vspace{-8pt}
\subsection{Apache TVM}

Apache TVM~\cite{tvm} is an end-to-end machine learning compiler framework for CPUs, GPUs, and specialized accelerators. It is actively used and supported by a wide community of developers and researchers worldwide. 
TVM allows compilation, optimization and deployment of DNNs in a variety of hardware acceleration devices. It also features a very powerful debugger, which allows exporting metadata regarding the model inference, such as layer activations, execution times per-model layer, as well as metadata about model structure and parameters, each exported in separate metadata files, while also providing a high-level API for their analysis.
We utilize the TVM Debugger in order to perform Layer Activation Analysis, as described in Section~\ref{layer-activations-analysis}.

\vspace{-5pt}
\section{Related Work}
To achieve model conversions between DL frameworks, a variety of tools exist, such as tf2onnx~\cite{tf2onnx}, tflite2onnx~\cite{tflite2onnx}, onnx2keras~\cite{onnx2keras}, onnx2torch~\cite{onnx2torch}, and native DL framework converters such as ONNX converter from PyTorch~\cite{pytorch} and \texttt{TFLiteConverter} from TFLite~\cite{tensorflow2015-whitepaper}. 
However, these tools come with no guarantees, and can potentially introduce faults in the converted models. To this direction however, limited research is conducted in the literature. Openja et al.~\cite{dlfconversionsstudy} published a study that emphasizes the challenges of the conversion process. Jajal et al.~\cite{InteroperabilityConverters} conducted a failure analysis on ONNX model converters, concluding that converters are responsible for about 75\% of the conversion defects, and that 33\% of reported failures are related to semantically incorrect models.
Liu et al.~\cite{MMdnn} propose MMdnn, a framework to automate the framework conversion process and mitigate a number of conversion challenges that result in conversion failures, such as unavailable operators, inconsistent tensor layouts, unsupported padding and incompatible argument types. 
MMdnn, however, does not consider faults related to configuration, such as input preprocessing, or faults related to additional important aspects such as layer parameters and computation graph that can also result in output incompatibilities during conversion. This is addressed by \tool.
It is also worth noting that MMdnn is currently in a deprecated state, with its last update on August 2020. In addition, our efforts with testing MMdnn was unsuccessful as it was built with older, deprecated versions of DL frameworks (e.g., PyTorch v0.4.0, while the current version is v2.2.0). We rigorously~\cite{louloudakis2023exploringeffectscomputationalparameter, louloudakis2022assessing, louloudakis2023deltann} explored the effects of the computational environment parameters on image recognition models, including deep learning (DL) framework conversions, and implemented a proof-of-concept~\cite{LouloudakisBuggyConversions} concerning conversion faults related to model weights and biases, which is the predecessor of \tool.

In an attempt to compare MMdnn with our work, we identified a potential overlap of only $15$ potential faults out of the $462$ we repaired with \tool, based on the results reported in~\cite{MMdnn}. 
Finally, MMdnn supports conversions over a restricted set of DL frameworks (primarily CNTK, CoreML, Keras, MXNet, PyTorch, TensorFlow, Caffe, and DarkNet) that is not currently maintained or updated while our approach is generalizable to model conversions between any pair of DL frameworks.

Automatic fault localization of DNNs is also explored in the literature. 
DeepFault~\cite{deepfault} focuses on a suspiciousness-oriented spectrum analysis algorithm to identify problematic model parts, and proposes a method for adversarial input generation. 
DeepLocalize~\cite{wardat2021deeplocalize} converts the model into an imperative representation and then performs dynamic analysis on its execution traces. 
In the context of fault localization of DL frameworks, CRADLE~\cite{pham2019cradle} performs an execution analysis on the model graph in order to detect faults, while LEMON~\cite{wang2020lemon} utilizes the metrics used by CRADLE and applies mutation testing to identify issues in the model. However, none of the aforementioned approaches focus on faults encountered during model conversions.

\vspace{-5pt}
\section{Methodology}
\label{methodology}

\begin{figure*}[!t]
 \centering
 \includegraphics[width=0.94\linewidth]{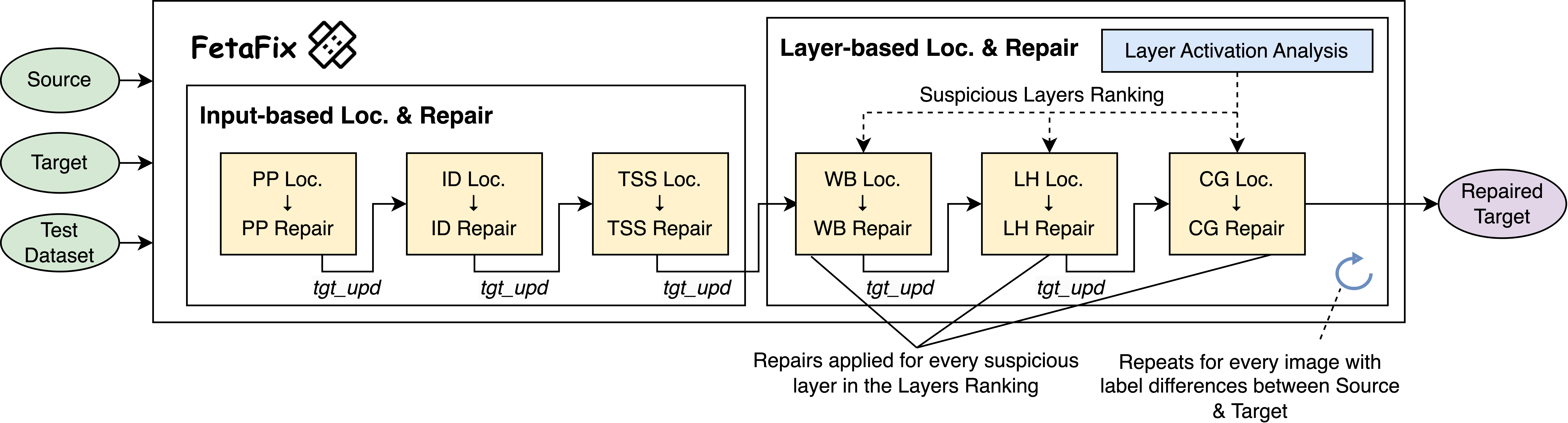}
\vspace{-10pt}
\caption{Fault localization and repair pipeline of \tool\ handling 6 fault types, three within Input-based category and three within Layer-based category.}
\label{fig:sysarch}
\vspace{-10pt}
\end{figure*}

\tool\ first takes as input the dataset, \source\ and \target\ models. It supports any image test dataset, requiring only the definition of a dataset path in its configuration.
It then localizes and repairs conversion errors that resulted in output label discrepancies between \source\ and \target. 
To compare the \source\ and \target\ model architectures, configuration and parameters we use their corresponding representation in the ONNX format which is a popular intermediate representation. We then import the \source-ONNX and \target-ONNX models to the TVM compiler framework to facilitate comparison. 
\tool\ comprises two main components: 1) A \emph{Fault Localization} component that identifies locations, parameters and properties in the \target\ model that may be potentially faulty; and 2) a \emph{Fault Repair} component with a defined set of strategies to fix the faulty locations and aspects identified through fault localization. The full fault localization and repair pipeline of \tool\ is presented in Figure~\ref{fig:sysarch}.


\vspace{-9pt}
\subsection{Fault Localization}
\label{fault-localization}

We surveyed $40$ posts and issues from StackOverflow, GitHub, and forums such as TensorFlow and NVIDIA discussions, where model conversion errors were raised, to understand the possible locations of errors. Although not exhaustive, we believe this sample to be representative of issues faced by developers, as (1) the issues we observed were among the most popular with regard to conversions and (2) the same or very similar issues appeared across different trackers and forums. Table~\ref{tab:faults-class} shows six different conversion fault types identified in these issues that were related to 1) Preprocessing (PP), (2) Input Dimensions (ID), (3) Tensor Shape and Structure (TSS), (4) Weights \& Biases (WB), (5) Layer Hyperparameters (LH), and (6) Computation Graph (CG). We also discovered many of these fault types examining the conversion errors reported in~\cite{louloudakis2023deltann}.

\begin{table}[t]
    \centering
    \vspace{3pt}
    \small
    \caption{Model conversion faults from StackOverflow, GitHub issues and other forums. Faults are classified as input-based and layer-based.}
    \vspace{-5pt}
    \label{tab:faults-class}
	\begin{tabular}{|l|l|l|V{5.5cm}|}
		\hline
		\# & Class & Fault & Sources \\
		\hline
		1 & Input & PP & \cite{stackoverflow_tflite_2023,stackoverflow_tf_incorrect_data_2023,stackoverflow_coreml_wrong_prediction_2023a,stackoverflow_tf_problem_or_wrong_2023,pytorch_forum_coreml_wrong_prediction_2020,tensorflow-lite-model-gives-very-different-accuracy-value-compared-to-python-mod} \\ 
		\hline
		2 & Input & ID & \cite{stackoverflow_dimension_mismatch_2023, 
			tensorflow_forum_dimension_mismatch_2023, 
			stackoverflow_valueerror_dimension_mismatch_1_2023, 
			stackoverflow_valueerror_dimension_mismatch_2_2023, 
			stackoverflow_dimension_mismatch_keras_to_onnx_2023, 
			github_pytorch2keras_dimension_issue_2023, 
			stackoverflow_model_dimension_error_2023, 
			github_onnx_tensorflow_conversion_issue_2023, 
			stackoverflow_strange_dimension_behaviour_2023} \\
		\hline
		3 & Input & TSS & \cite{github_onnx_keras_issue572_2023,
			github_pytorch2keras_issue31_2023,
			github_pytorch2keras_issue78_2023,
			github_pytorch2keras_issue76_2023,
			github_microsoft_MMdnn_issue831_2023,
			github_onnx_tensorflow_issue782_2023} \\
		\hline
  		4 & Layer & WB  & \cite{github_pytorch2keras_issue127_2023,
			github_pytorch2keras_issue124_2023,
			github_microsoft_MMdnn_issue823_2023,
			github_tensorflow_issue35194_2023,
			tensorflow_forum_accuracy_drop_2023,
			stackoverflow_tflite_conversion_changes_weights_2023a,
			stackoverflow_tflite_model_overflows_on_gpu_2023,
			stackoverflow_accuracy_drop_tensorflow_to_tflite_2023,
			tensorflow_forum_model_accuracy_loss_2023,
			github_tensorflow_issue31359_2023,
			github_tensorflow_issue31205_2023,
			github_microsoft_MMdnn_issue297_2023} \\
            \hline
		5 & Layer & LH & \cite{github_onnx2keras_issue135_2023,
			github_sagemaker_python_sdk_issue613_2023, nvidia_forum_strides_problem_2021} \\
		\hline
		6 & Layer  & CG & \cite{github_pytorch2keras_issue135_2023,
			github_onnx_tensorflow_issue2246_2023, github_tensorflow_onnx_issue1203_2023, stackoverflow_poor_tflite_accuracy_android_2023} \\
		\hline
	\end{tabular}
\vspace{-15pt}
\end{table}

As seen in Figure~\ref{fig:sysarch}, our fault localization approach starts by localizing the first three fault types: PP, ID, and TSS. These three fault types pertain to the input, stemming from differences between \source\ and \target\ with respect to the preprocessing used, or changes in input dimensions or operations on the input. We refer to these fault types as \emph{Input-based}. 
After attempting to localize input-based fault types, our approach proceeds with the remaining three fault types in Table~\ref{tab:faults-class} that belong to the \emph{Layer-based} category, as it stems from differences in layer hyperparameters, weights and biases, and differences in computation graph.

\vspace{3pt}
\noindent 
\textbf{Comparisons for Input-Based Fault Types} \\
\vspace{2pt}
We examine and contrast input-related settings between the \source\ and \target\ models in the fault types below. 

\vspace{-7pt}
\subsubsection{Preprocessing (PP)}

For some instances, differences in preprocessing settings between \source\ and \target\ can result in output label differences. We explore the effect of different preprocessing configurations for \target\ using the default preprocessing setting from the official repositories of the DL frameworks both \source\ and \target. 
For example, in the case of converting MobileNetV2 from PyTorch to Keras, \tool\ examines the performance of the converted \target, using both official PyTorch and Keras preprocessing settings for MobileNetV2.

\vspace{-7pt}
\subsubsection{Input Dimensions (ID)}
\label{input-comparison}

Islam et al.~\cite{repairingDNNs} identified that changes in layer input dimensions can affect the performance of a DNN model. We also empirically identified that model conversion tools accept misconfigurations of \target\ model input dimensions without notifying the user of the mismatch between \source\ and \target\ input dimensions that may potentially result in output label discrepancies. 
We explore problems arising from such misconfigurations in this fault type.
For example, we inspected the conversion of ResNet101 from PyTorch to TFLite, modifying the input dimensions from \texttt{(1, 3, 224, 224)} used in \source\ to \texttt{(1, 3, 299, 299)} used in \target. The conversion process completed without errors and warnings. However, the model presented a significant output label discrepancy between \source\ and \target\ ($37$\%). In our fault localization approach, we check if input dimensions between \source\ and \target\ match, and report any mismatches to the fault repair component. 

\vspace{-7pt}
\subsubsection{Tensor Shape and Structure (TSS)}
\label{tss-localization}

Another potential error identified by Islam et al.~\cite{repairingDNNs} relates to tensor shapes and changes in model structure. We consider cases where an erroneous tensor shape can affect the correctness of the model conversion where \source\ and \target\ models have differences in output labels. 
Such a scenario may be observed when a transpose operation is introduced by a converter tool to modify the model input. This is typically manifested as a \texttt{Transpose} node right after the input node in the computation graph. In an erroneous conversion scenario, the \texttt{Transpose} node might generate a tensor with correct shape, but incorrect structure. 
To illustrate this, consider the example of a \source\ input with the shape \texttt{(1, 224, 224, 3)}. The \target\ model on the other hand has a shape of \texttt{(1, 3, 224, 224)} after applying a \texttt{Transpose} node that modifies the input of \target, by either applying \texttt{Transpose(tensor, [0, 2, 3, 1])}, or \texttt{Transpose(tensor, [0, 3, 2, 1])}, depending on the implementation. 
The modified shape is used in subsequent nodes in the computation graph. Owing to the ambiguity in dimensions where both transpose operations result in a tensor with correct shape expected by the subsequent layers in the \target\ model, it can result in model conversion errors as one of them has an incorrect structure. 
To localize faults like this, \tool\ checks if there is a difference in model input shape between \source\ and \target. For cases presenting differences, \tool\ further examines if any of the input-related subsequent layers contains a shape transformation operation (like \texttt{Transpose}) and considers such scenarios as potential sources of faults, conservatively.

Potential faults related to tensor structure may also happen when a layer tensor in the \target\ model is incorrectly transformed by the converter and is followed by a flattening or a reshaping operation that loses the original structure information, making the difference with \source\ model non-obvious.
In particular, if a tensor has a different shape between \source\ and \target\ (i.e., due to erroneous conversion) but the total number of elements is the same across tensors, a \texttt{Flatten} or a \texttt{Reshape} operation on such a tensor will make it impossible to detect the mismatch between \source\ and \target, as the structure information is lost.
However, the order of elements between the corresponding \source\ and \target\ tensors will be different after the \texttt{Flatten} or a \texttt{Reshape} operation, which may result in output label discrepancies.
For instance, if a layer has shape \texttt{(1, 3, 32, 32)} on \source\, but was converted to \texttt{(1, 32, 32, 3)} on \target, and is followed by a \texttt{Flatten} node in both models, the result will be tensors with \texttt{(3072)} elements in both models. We localize this fault by examining the tensor shape of the dominator nodes preceding any \texttt{Flatten} or \texttt{Reshape} nodes, and check if it matches with the corresponding nodes in the \source\ model.\\

\begin{figure}[t]
 \centering \includegraphics[width=\linewidth]{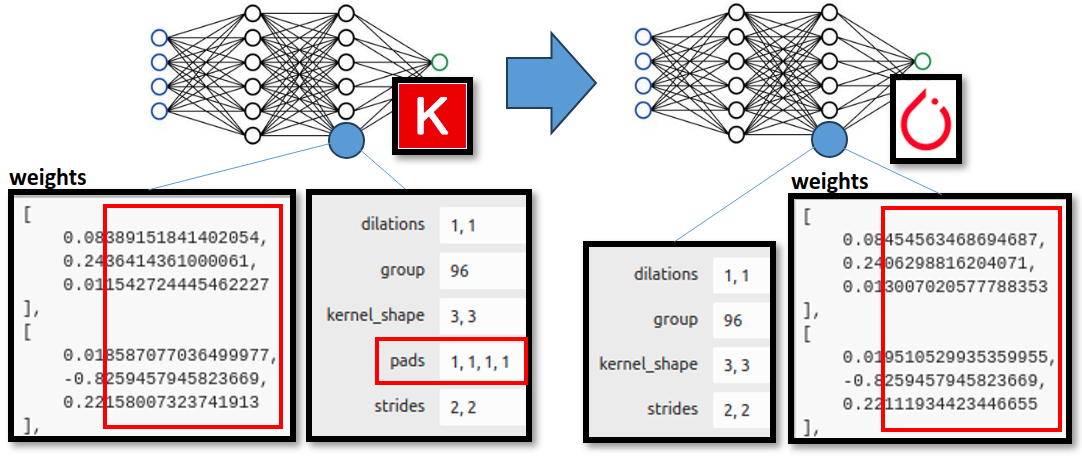}
 \vspace{-15pt}
 \caption{Indicative example of differences in layer weights and hyperparameters, introduced in the model conversion process.}
 \vspace{3pt}
 \label{fig:discrepancies-illustration}
\end{figure}

\vspace{-8pt}
\noindent \textbf{Comparisons for Layer-Based Fault Types} \\
\vspace{2pt}
For this category of faults, we examine and compare layer activations between \source\ and \target\ models. For layers where differences in activations are outside of an  expected range of differences (as described in Section~\ref{layer-activations-analysis}), we mark those layers as suspicious and rank them so that we examine the most suspicious layer first for fault types CG, LH, and WB. 
We then iteratively proceed down this ranked list of suspicious layers for inspecting the three layer-based fault types, CG, LH and WB, discussed below. 
We start by describing our approach for producing a ranked list of suspicious layers in layer activation analysis followed by our fault localization approach for CG, LH, and WB with the ranked suspicious layers. 
An example instance of the differences between \source\ and \target\ in weights, biases, and hyperparameters that \tool\ attempts to localize as potential faults is shown in Figure~\ref{fig:discrepancies-illustration}. We find, for a core computational layer under inspection, that the weights present small differences between \source\ and \target. Additionally, the \texttt{pads} hyperparameter that is present in \source, is missing on \target\ in Figure~\ref{fig:discrepancies-illustration}.


\vspace{-8pt}
\subsubsection{Layer Activation Analysis}
\label{layer-activations-analysis}

We compare the activations for each convolutional layer between \source\ and \target\ models, as a four-step process.
(1) We extract two equally sized sets of images from the test dataset: one for \emph{similar} images (images that produce the same top label prediction across \source\ and \target), and one for \emph{dissimilar} images (images with different labels between \source\ and \target).
We then extract the layer activation tensors of both sets by loading the \source\ and \target\ ONNX models to TVM and performing inference using the \textit{TVM Debugger}~\cite{tvm_debugger}. 
The TVM debugger generates the complete model graph and parameters at compile time, while also gathering layer activations for each input.
Using these activations and the \source\ ONNX model, we then utilize the \texttt{LAYER\_ANALYSIS} function, shown in Algorithm~\ref{alg:layer-analysis}.
This function takes as input the original source ONNX model (used to iterate the matching layer activations for both \source and \target), as well as the activations for the sets of similar and dissimilar images for both \source and \target (contained in parameters \texttt{sim\_acts} and \texttt{diss\_acts}). It also utilizes the \texttt{List} module, responsible for list length check and sorting operations, and the Kruskal-Wallis parametric test (\texttt{KWTest}).
(2) To identify when activation mismatches between \source\ and \target\ indicate a potential problem, we first extract the differences in activations between \source\ and \target, per element within each layer, for the set of similar images and store these differences in a set, \texttt{sim\_diff} in lines $8$ and $9$ of Algorithm~\ref{alg:layer-analysis}.
We treat differences in activations between \source\ and \target\ for these similar images as acceptable differences and we use it to estimate the expected difference in distribution of activations for correct conversions.
(3) Next, for the set of dissimilar images, we compute the set of dissimilar differences, \texttt{diss\_diff} (lines $11$ and $12$), per tensor element per layer. We then check if the difference in activations per tensor element per layer belongs to the corresponding expected distribution of activation differences (from the previous step).
To do this, we use  the \textit{Kruskal-Wallis non-parametric statistical test} (line $13$)~\cite{kruskallwallistestmethodology}\footnote{The distribution of activation differences for correct conversion instances did not follow a normal distribution, so we chose the non-parametric Kruskal-Wallis test.} to determine if \texttt{sim\_diff} and \texttt{diss\_diff} belong to the same distribution (using a standard 5\% significance level) (line $17$).
(4) From step 3,  we obtain the tensor elements whose difference in activations for \texttt{sim\_diff} and \texttt{diss\_diff} did not belong to the same distribution, for each layer between \source\ and \target. These are considered as potentially \emph{problematic} elements that may result in model conversion errors. We count the number of problematic tensor elements for each model layer (line $18$) and rank the layers in descending order of number of problematic elements (line $20$). This ranking of suspicious layers is the basis for fault localization and repair of fault types CG, LH, and WB. We only consider convolutional layers as these are the core computational layers for a Convolutional Neural Network (CNN). However, this can be modified to consider other layer types as well.

\vspace{-5pt}
\setlength{\textfloatsep}{0pt}
\begin{algorithm}[t]
\caption{Layer Analysis in \tool}
\label{alg:layer-analysis}
\begin{algorithmic}[1]
\small
\Require{List, KWTest, model, sim\_acts, diss\_acts}
\Procedure{layer\_analysis}{model, sim\_acts, diss\_acts}
    \For{layer \textbf{in} model.layers}
        \State layer.prob\_elems $\gets$ 0
        \For{element \textbf{in} layer.elements}
            \State sim\_diff, diss\_diff $\gets$ [], []
            \State // Gather element-wise activations.
            \For{sim \textbf{in} sim\_acts}
                \State sim\_diff $\gets$ sim\_diff $\cup$
                \State\hspace{4.9em}(sim.tgt[element] - sim.src[element])
            \EndFor
            \For{diss \textbf{in} diss\_acts}
                \State diss\_diff $\gets$ diss\_diff $\cup$
                \State\hspace{4.9em}(diss.tgt[element] - diss.src[element])
            \EndFor

            \State kw\_out $\gets$ \Call{KWTest}{sim\_diff, diss\_diff}
            \State // If the null hypothesis of
            \State //  same distribution is rejected,
            \State // then the element counts as problematic.
            \If{kw\_out.p\_value $<$ 0.05}
                \State layer.prob\_elems$++$
            \EndIf
        \EndFor
    \EndFor
    \State // Sort layers in descending order of problematic elements.
    \State \Return \Call{sort}{model.layers,\State
    \hspace*{5.3em}(l1, l2) $\Rightarrow$ l1.prob\_elems $-$ l2.prob\_elems}

\EndProcedure

\end{algorithmic}
\end{algorithm}

\subsubsection{Weights and Biases (WB)}
\label{weights-biases-localization}

We extract the weight and bias tensors for each of the convolutional layers. Starting from the most suspicious layer in the ranked list, we compare the corresponding \source\ and \target\ tensors element-wise for each of the layers. We flatten the tensors to facilitate comparison. For correct model conversions, we expect the weights and biases for the \source\ and \target\ model layers to match exactly. Values for model weights and biases are only affected during model training or optimization, neither of which is part of model conversion. 
In addition to convolutions, we also perform comparison of weights and biases in neighboring layers, e.g., batch normalization nodes preceding and/or succeeding the convolutional nodes).
Consequently, differences identified during the comparison of weights and bias tensors for layers across \source\ and \target\ point to potential problems in those layers. We output all the layers that have differences identified in weights and bias tensors between \source\ and \target\, and pass it to the repair component.

\vspace{-5pt}
\subsubsection{Layer Hyperparameters (LH)}
\label{hyperparameters-localization}

Incorrectly converted hyperparameters are another potential source of error. For example, with a convolutional layer, we expect that the padding, strides, dilation, and other configurations to remain unchanged during model conversion.
Starting from the most suspicious layer in the ranked list, we compare layer hyperparameters and attributes -- (\texttt{pads, strides, kernel\_shape, dilations, epsilon, min, max, axis}) between \source\ and \target. If we find any differences, we mark them as potential faults.
In particular, we observe $3$ types of differences: (1) a hyperparameter is present in \source\ but not in \target; (2) a hyperparameter is present in \target\ but not in \source\ (e.g., added by the converter tool); and (3) a hyperparameter is present both in \source\ and \target but their values are different.

\vspace{-5pt}
\subsubsection{Computation Graph (CG)}
\label{graph-localization}

In our experiments, we found that output label discrepancies can sometimes arise from differences in model graph between \source\ and \target\ models. We examine differences in graph nodes. 
To make the analysis tractable, we start from the most suspicious layer (based on the ranking from layer activation analysis) in \target\ and extract the subgraph between this layer and its dominator in the \target\ model graph. 
We do the same with the \source\ graph starting from the corresponding \source\ layer node and its corresponding dominator. We perform a depth first traversal of the two subgraphs and compare their nodes. Any differences are marked as a potential fault. We repeat this for any other suspicious layers in the ranked list.

\vspace{-8pt}
\subsection{Fault Repair}
\label{fault-repair}

We implemented a number of strategies to repair the potential faults identified by the fault localization component. We discuss the strategies to fix each of the issues identified in fault localization below. 
As seen in Figure~\ref{fig:sysarch}, for layer-based repairs to CG, LH, and WB, the fault repair component will start from the most suspicious layer in the layer ranking provided by the layer activation analysis (within fault localization) and apply fixes sequentially and iteratively to layers in that ranking. Each of the fixes to \target\ produces an updated version. The updated \target\ will only be accepted if the output label discrepancy with respect to \source\ reduces after the repair (compared to previous target version), and this is then used as the current \target\ for the next repair strategy or suspicious layer in the list. 
On the other hand, for PP, ID, TSS, the repair actions are layer-independent, and do not need to consider the layer ranking. The repair strategies are applied sequentially with PP applied first, followed by ID and then TSS. As with layer-based strategy, the fix is only accepted if the output label discrepancy is reduced. Finally, both input-based and layer-based strategies are applied to every image in the dataset that had different labels between \source\ and \target\ until they are all fixed or the algorithm times out.

\vspace{5pt}
\noindent \textbf{Input-based Repair Strategies}
\vspace{-7pt}
\subsubsection{Preprocessing Differences (PP) Repair}
    We experiment running the \target\ model with both the \target\ and \source\ preprocessing settings to check if the label discrepancies are resolved.
    \vspace{-5pt}
    \subsubsection{Input Dimensions (ID) Repair} 
    Following fault localization, if a difference is detected between \source\ and \target\ input dimensions, \tool\ automatically fixes \target\ to match with the input dimensions from \source. We observed four erroneous model conversions in our experiment with incorrect input dimensions, with two of them requiring further actions in model nodes for tensor shape and structure (discussed below).
    \vspace{-5pt}
    \subsubsection{Tensor Shape and Structure (TSS) Repair} 
    Following fault localization, as described in Section~\ref{tss-localization}, if the input is different between \target\ and \source\ in a problematic conversion, and input is succeeded by a \texttt{Transpose} node, then \tool\ attempts to apply a fix by removing that node, in addition to adjusting the \target\ input to match \source. Furthermore, if \tool\ detects any nodes that are related to input processing and normalization (e.g., InceptionV3 architecture contains \texttt{Gather}, \texttt{Unsqueeze}, \texttt{Add}, and \texttt{Mul} nodes before the added \texttt{Transpose}), then \tool\ adjusts the attributes of these nodes in order to be compatible to the \source\ input dimension. In cases where \tool\ detects the presence of \texttt{Flatten} or \texttt{Reshape} nodes that alter the shape of tensors, then it attempts to repair this by inserting a \texttt{Transpose} node right before \texttt{Flatten} or \texttt{Reshape} to match tensor shape with \source.

\vspace{5pt}
\noindent \textbf{Layer-based Repair Strategies}
\vspace{-7pt}
    \subsubsection{Weights and Biases (WB) Repair}
    If a mismatch is found between any weight or bias tensor across model layers by the fault localization component, then \tool\ replaces the values from the corresponding tensor in \source\ to \target. As mentioned earlier, \tool\ performs the repair starting from the most suspicious layer in the ranked list.
    \vspace{-5pt}
    \subsubsection{Layer Hyperparameters (LH) Repair} 
    Upon marking a layer hyperparameter case as potentially problematic, as described in Section~\ref{hyperparameters-localization}, \tool\ attempts to fix \target\ by applying the corresponding setting of \source. In particular, if (1) a node is present in \source\ but is missing from the computation graph of \target, \tool\ will update the corresponding node including the \source\ hyperparameter to \target. If (2) a hyperparameter is added in \target\ but is missing from \source, \tool\ will remove it from the target node. And finally, if (3) there is a value mismatch between \source\ and \target\ for the hyperparameter, \tool\ will update \target\ in accordance to the source value.
    \vspace{-5pt}
    \subsubsection{Computation Graph (CG) Repair} 
    In accordance to Section~\ref{graph-localization}, if a difference in subgraphs
    is detected between \source\ and \target, \tool\ attempts to repair it by replacing the \target\ subgraph nodes with \source\ subgraph nodes, preserving all the node properties and ensuring all weights, biases and hyperparameters are correctly updated in the \target\ graph.
\\
Regarding the order in which our repair strategies are applied, we first applied the ones that were input-based, namely PP, ID, and TSS, considering that issues introduced at this early processing stage can lead to problematic behavior accumulating and spreading throughout the model. We do PP first as changes to preprocessing may fix errors observed downstream. 
We then examine ID, with TSS being an additional restriction to ID mismatches. We then apply repairs that are ``layer-based'' -- WB, LH, and CG. Among the ones with layer-based effects, we first perform repairs related to model weights, biases and hyperparameters, and then those related to computation graph modifications. Note that some graph transformations may also be related to hyperparameter changes, e.g., removing a \target\ \texttt{Pad} node to adapt to \source\ structure requires the addition of a \texttt{padding} hyperparameter to the next convolutional layer. 

We implemented \tool\ in Python. We used Apache TVM Debugger for layer activations analysis.
The debugger allows the extraction of model graph structure and parameters in metadata files, and a log of the layer activations throughout the model structure.
In order to implement the repair strategies, we used the ONNX API to modify \target.


\vspace{-5pt}
\section{Experiments}
We consider three widely used image recognition models of various sizes: MobileNetV2~\cite{mobilenetv2}, ResNet101V2~\cite{resnet, resnetv2}, and InceptionV3~\cite{inceptionv3}.
We use models pre-trained on ImageNet~\cite{deng2009imagenet} using native model definitions and pre-trained parameters/weights sourced from four different DL frameworks' repositories: \textit{Keras}~\cite{chollet2015keras}, \textit{PyTorch}~\cite{pytorch},
\textit{TensorFlow(TF)}~\cite{tensorflow2015-whitepaper}, and \textit{TFLite}~\cite{tensorflow2015-whitepaper}.
For our experiments, we used two datasets: the ILSVRC2017~\cite{ILSVRC17} object detection test dataset consisting of $5500$ images, and the ImageNetV2 dataset~\cite{imagenetv2} consisting of $10,000$ images. 
We used the ILSVRC2017 dataset for detecting and repairing errors in model conversion. We then assessed the correctness of the repaired models on ImageNetV2, utilizing it as an independent dataset, to verify our results and avoid dataset overfitting. We used the problematic conversion cases identified in~\cite{louloudakis2023deltann} as our candidates for fault localization and repair. Both datasets contain real, non-synthetic samples and they are widely utilized by developers and researchers, while remaining tractable in size. 
We focused on cases where the conversion process introduced label discrepancies between \source\ and \target, and there were $15$ such erroneous conversions, as seen in Figure~\ref{fig:conversionsorig}. Overall, we selected models in the domain of computer vision for the following reasons: (1) To test our repair approach against the conversion faults we detected in the models we examined in our previous work~\cite{louloudakis2023deltann}; and (2) since our experiments involved 15 model conversion cases with numerous iterations and analysis within each, we chose image recognition models as they are more tractable and need less resources than other DL architectures, like transformers, for our experiments. However, our approach can generalize easily to any DL framework or conversion tool. It can be applied to any DL model architecture, including non-CNN models, only requiring the user to specify the core computational layers to be used in the comparison.

We evaluate the following research questions in our experiments: 
\begin{description}
\item[RQ1:] How effective is \tool\ at localizing faults in the $15$ erroneous model conversions?
\item[RQ2:] How effective is \tool\ at repairing the faults in the $15$ erroneous model conversions? We evaluate which of the repair strategies are used most frequently in model fixes. 
\end{description}

To check the robustness of \tool\ with respect to devices, we ran our experiments on two devices of varying capabilities, an Intel-based server featuring a high-end \texttt{Nvidia Tesla K40c (GK11BGL) GPU}, and a Laptop featuring a low-end \texttt{Intel(R) GEN9 HD Graphics NEO}. Our experiment results remained the same on both devices. 

\vspace{-7pt}
\subsection{Fault Analysis Setup}

Within layer activation analysis, discussed in fault localization for layer-based types, we determine the expected distribution of differences in activations per layer between \source\ and \target\ for images where the output label is the same using a random sample of up to $100$ images that produced similar results. We then compare the activations for dissimilar images against this expected distribution to come up with a ranking of suspicious layers. We repeat this process $3$ times with three different random samples of similar images.


\vspace{-7pt}
\subsection{Fault Repair Setup}

Once we extracted the order of layers from the fault analysis, we utilized the fault repair component of \tool, applying the strategies in a greedy approach, as described in Section~\ref{fault-repair}. 
We set a time limit of 2 hours for each conversion case. We also set a limit of three unsuccessful repair iterations, when repairs did not reduce output label discrepancy, as a stopping condition. This is to help make the experiments tractable. These settings can be customized as needed in our tool.

\vspace{-5pt}
\section{Results}
Overall, \tool\ was able to localize $755$ differences in the structure and properties between \source\ and \target\ variants across all $15$ erroneous conversion cases. We identify these differences as potential faults - given that (1) they might not affect model correctness and their impact needs further examination, and (2) many of those differences might have common roots of cause in properties, structure, as well as the converter tools, affecting \target\ models during the conversion process. \tool\ attempts to distinguish which of those differences had an actual impact to the model correctness and only repair those, while discarding the rest of them.
Furthermore, an overview of the potential faults distribution can be found in Table~\ref{tab:faultsdistribution}. 
From these $755$ cases \tool\ identified as potential bugs, it correctly distinguished as actually problematic and repaired $462$ of them ($61.1\%$) --- completely repairing $12$ out of $15$ problematic model conversions, while significantly improving the performance of $2$ more, with the most distinctive case being improved from $48.9$\% output label discrepancy to $0.33$\%. Our system was unable to reduce the percentage further, as none of our strategies affected the model beyond this point.


\vspace{-3pt}
\subsection{Fault Localization}

The faults localized were distributed across all six types within input-based and layer-based fault categories. 
Of the $755$ faults localized, $462$ of them ($61.1$\% of total cases) resulted in reducing output label discrepancy between \source\ and \target\ when repaired. For the remaining localized faults, FetaFix attempted to repair them by applying the respective repair strategy, but did not result in any improvement. Furthermore, they are considered as false positives detected by the fault localization component and are automatically discarded from the repair process. We demonstrate the distribution of faults detected and fixed across the different types in Table~\ref{tab:faultsdistribution} and discuss each type in the context of fault localization next.

Overall, the faults we detected are new, as we detected no reported issues with relation to them in the respective converter repositories. We aim reporting the issues to the respective developers of the conversion tools.

\begin{table}[t]
    \centering
    \vspace{3pt}
    \caption{Number of faults localized versus repaired (\#Localisations/\#Repairs) per model conversion case.}
    \vspace{-8pt}
    \label{tab:faultsdistribution}
    \small
    \begin{tabular}{V{0.19cm}|l|l|l|l|l|l|l}
    \hline
        ~ & ~ & \multicolumn{6}{|c}{\textbf{\#Localisations/\#Repairs}} \\
        \hline
        ~ & ~ & \multicolumn{3}{c|}{\textbf{Input-Based}} & \multicolumn{3}{c}{\textbf{Layer-Based}}\\ 
        \hline
        \textbf{\#} & \textbf{MobileNetV2} & PP & ID & TSS & WB & LH & CG \\ \hline
        1 & Keras-to-Torch & \textbf{1/1} & 0/0 & 0/0 & 0/0 & \textbf{6/0} & 0/0 \\ \hline
        2 & Keras-to-TFLite & \textbf{1/1} & 0/0 & 0/0 & \textbf{96/96} & 0/0 & 0/0 \\ \hline
        3 & Torch-to-TF & \textbf{1/1} & 0/0 & 0/0 & 0/0 & \textbf{6/0} & 0/0 \\ \hline
        4 & Torch-to-Keras & \textbf{1/1} & \textbf{1/1} & \textbf{1/1} & 0/0 & \textbf{39/0} & 0/0 \\ \hline
        5 & TF-to-TFLite & 0/0 & 0/0 & 0/0 & \textbf{68/66} & 0/0 & 0/0 \\ \hline
        6 & TF-to-Keras & \textbf{1/1} & 0/0 & 0/0 & 0/0 & \textbf{40/0} & \textbf{4/0} \\ \hline
        ~ & ~ & ~ & ~ & ~ & ~ & ~ & ~ \\ \hline
        ~ & \textbf{ResNet101} & ~ & ~ & ~ & ~ & ~ & ~ \\ \hline
        7 & Keras-to-Torch & \textbf{1/1} & 0/0 & 0/0 & 0/0 & \textbf{72/0} & \textbf{2/0} \\ \hline
        8 & Keras-to-TFLite & 0/0 & 0/0 & 0/0 & \textbf{180/174} & \textbf{33/0} & \textbf{37/0} \\ \hline
        9 & Torch-to-TFLite & \textbf{1/1} & \textbf{1/1} & 0/0 & 0/0 & 0/0 & 0/0 \\ \hline
        10 & Torch-to-TF & \textbf{1/1} & 0/0 & 0/0 & 0/0 & 0/0 & 0/0 \\ \hline
        11 & TFLite-to-Keras & 0/0 & 0/0 & 0/0 & 0/0 & \textbf{4/4} & \textbf{3/3} \\ \hline
        12 & TF-to-TFLite & 0/0 & 0/0 & 0/0 & 0/0 & 0/0 & 0/0 \\ \hline
        13 & TF-to-Keras & 0/0 & \textbf{1/1} & \textbf{1/1} & 0/0 & \textbf{1/1} & \textbf{1/1} \\ \hline
        ~ & ~ & ~ & ~ & ~ & ~ & ~ & ~ \\ \hline
        ~ & \textbf{InceptionV3} & ~ & ~ & ~ & ~ & ~ & ~ \\ \hline
        14 & Torch-to-Keras & \textbf{1/1} & \textbf{1/1} & \textbf{1/1} & 0/0 & \textbf{54/7} & 0/0 \\ \hline
        15 & TF-to-TFLite & 0/0 & 0/0 & 0/0 & \textbf{94/94} & 0/0 & 0/0 \\ \hline
        ~ & \textbf{Total} & \textbf{9/9} & \textbf{4/4} & \textbf{3/3} & \textbf{438/430} & \textbf{254/12} & \textbf{47/4} \\ \hline
    \end{tabular}
    \vspace{7pt}
\end{table}

\noindent \textbf{Preprocessing (PP):}
We detected $9$ cases in which the choice of preprocessing between \source\ and \target\ configurations was a decisive factor for the model performance, while $5$ cases presented the same results across \source\ and \target\ configurations.
All models presented severe performance degradation in output correctness if no preprocessing option was selected.

\noindent \textbf{Input Dimensions (ID):}
\tool\ identified four model cases where the input dimensions were different between \source\ and \target. We identified that input dimension errors had two primary causes: (1) erroneous configuration from the user, which neither the model converters nor the model itself raised as warnings or errors; (2) implicit changes in model input due to the converter implementation. 

We observed two cases, $9$ and $13$ in Table~\ref{tab:faultsdistribution}, where ID fault is a result of the first cause. Regarding the second cause, we detected two cases where the converter changed the model input. In addition to ID faults, we also detected differences in tensor shape and structure that we discuss as part of the next category.

\noindent \textbf{Tensor Shape and Structure (TSS):}
\tool\ detected three cases with potential faults in the TSS category.
The model converter introduced a \texttt{Transpose} node in these two cases to address the difference in ID, so that subsequent nodes use inputs with the correct dimension. However, in these two cases achieving the correct input dimension by using the \texttt{Transpose} node was not adequate. This occurred due to similarly sized dimensions within tensors, like the third and fourth dimension within an example tensor with shape \texttt{(1, 3, 224, 224)}, where it is possible to incorrectly transpose dimensions of the same size that result in a different structure from what is intended.

In addition, in some model architectures (e.g., InceptionV3) the \texttt{Transpose} node is placed after a series of nodes that normalize data based on a specific input dimension. For instance, in InceptionV3, the \texttt{Transpose} node is placed after a series of \texttt{Gather}, \texttt{Unsqueeze}, \texttt{Mul}, and \texttt{Add} nodes. \tool\ detects these differences across model graph and reports them during fault localization.

\noindent \textbf{Weights and Biases (WB):}
We observed $438$ cases of tensor value differences in weights and biases between \source\ and \target, with ResNet101 converted from \texttt{Keras} to \texttt{TFLite} containing $177$ of them. 

\noindent \textbf{Layer Hyperparameters (LH):}
\tool\ detected $254$ potential faults across $12$ model conversion cases, with most of them concerning padding hyperparameters. \tool\ also identified some cases where a potential hyperparameter fault was associated with a potential computation graph (CG) fault - for instance, specific convolutional layers presented potential faults related with padding hyperparameters and \texttt{Pad} nodes that directly preceded these convolutional layers.

\noindent \textbf{Computation Graph (CG):}
\tool~identified $47$ cases across $4$ conversion cases in which parts of the model computation graph presented discrepancies across \source\ and \target\ variants. All the cases observed were associated primarily with three operations: (1) setting a padding around a tensor, by utilizing \texttt{Pad} nodes; (2) performing batch normalization to a tensor by applying an element-by-element constant addition, and scaling using a \texttt{BatchNorm} node; and (3) redundant output nodes erroneously placed by the converter. From those $47$ cases, \tool~successfully identified $4$ of them as actual bugs and repaired them, playing a crucial role for the repair of cases \#$11$ and $13$.


\vspace{-15pt}
\subsection{Fault Repair}

From the $755$ potential faults localized and reported in Table~\ref{tab:results-repair}, \tool\ successfully repaired $462$ of them across the $15$ erroneous mode conversion scenarios. In $12$ out of the $15$ cases, \tool\ was able to repair the model completely, starting from up to $100\%$ initial output label discrepancy. In two other cases, \tool\ was effective reducing the output label difference between \source\ and \target\ to less than $1.5$\%.
The only case that \tool\ was unable to repair was because the \source~ and \target~ models could not compile in the same opset, therefore the graph representation of the model was different, deeming it unable to detect potential model discrepancies. 
By performing manual inspection across both \source\ and \target, we observed that \target\ was representing the weights and biases of each layer in separate nodes, while on \source\ weights and biases were associated as layer inputs. \tool\ is implemented to match weights and biases represented as inputs to each layer, and therefore was unable to repair this scenario.
We discuss fault repair with respect to each of the fault types below. \\


\vspace{-5pt}
\noindent \textbf{PP:} As shown in Table~\ref{tab:results-repair}, \tool\ was able to repair cases \#$1$, $3$, $6$, $7$, $9$  and $10$, by determining and setting the correct preprocessing configuration. Given that the preprocessing is not a standard procedure across models, we identified that for these cases using the \source\ preprocessing instead of the one officially provided for the \target\ library fixed the result completely.  
Cases \#$2$, $4$, $9$, and $14$ were also affected partially by preprocessing selection, however additional steps were needed in order to repair them. The additional steps are explained in more detail in the categories below.

\noindent \textbf{ID:} Among the $4$ faults localized as ID, \tool\ repaired one case completely by solely adjusting the input dimensions of \target\ to match \source.
For the remaining three cases, additional actions had to be performed, involving repair tensor shape and structure, as well as computation graph repair, as described in the next categories.

\noindent \textbf{TSS:} \tool\ was able to repair three cases related to problematic tensor shape and structure. In particular, cases $4$, $13$, and $14$ were related to problematic layers right after input. 
All three cases had different input dimensions between \source\ and \target, and the conversion tool had inserted additional \texttt{Transpose} nodes in order to adjust the input tensor shape so that it would be aligned with subsequent nodes. However, the transformation was done improperly, leading to severe performance degradation of up to $68.21$\%. We identified all cases to be related to the \texttt{onnxmltools}~\cite{onnxmltools} converter (part of the native ONNX codebase), when converting Keras to ONNX.
\tool\ automatically identified the presence of a redundant \texttt{Transpose} node following the  input node, and repaired it by (1) adjusting the \target\ model input dimensions so that they correspond to those of \source, (2) altering its tensor transposition order (permutation) property to align with the new input dimensions, and (3) updating the properties of nodes connected to \texttt{Transpose} to ensure proper tensor handling, reflecting the change in \texttt{Transpose}. This modification in particular was vital to case \#$14$ where \tool\ detected additional nodes (\texttt{Gather}, \texttt{Unsqueeze}, \texttt{Add} and \texttt{Mul}) preceding \texttt{Transpose} that apply input normalization. As part of the repair process, \tool\ updated the \texttt{axes}  properties of \texttt{Gather} and \texttt{Unsqueeze} nodes to align with the updated input dimension and the \texttt{Transpose} permutation in \target.

\begin{figure}[t]
     \centering
     \vspace{5pt}
     \includegraphics[width=\linewidth]{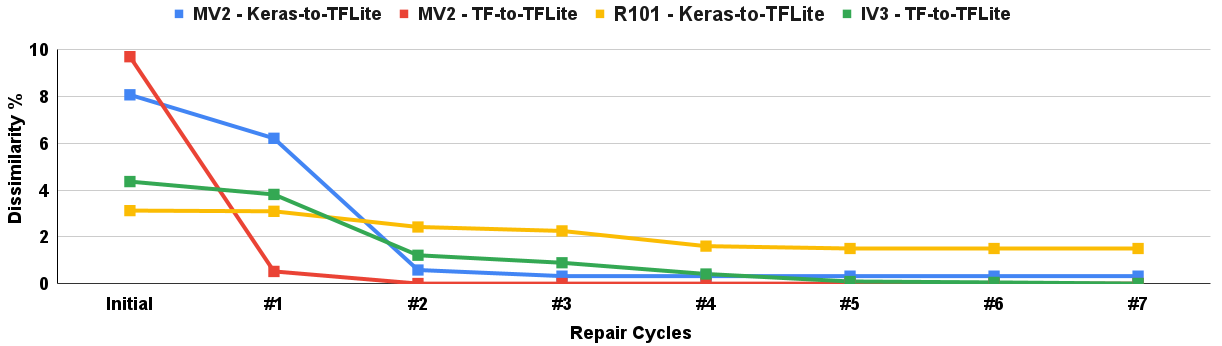}
     \vspace{-20pt}
     \caption{Model conversion cases that required an iterative weights and biases repair strategy with the percentage output label dissimilarity shown after each repair cycle (following preprocessing repair). }
     \label{fig:iter-wb-cases}
     \vspace{5pt}
\end{figure}

\noindent \textbf{WB:} Cases \#$2$, $5$, $8$, and $15$ were related to parameter precision discrepancies between \source\ and \target. Although the discrepancies were negligible, the model behavior deviated across the cases from $3.12$-$9.7$\%. \tool\ performed multiple iterations and managed to completely fix cases \#$5$ and $15$ while also significantly improving cases $2$ and $8$ by replacing the weights of \target\ with those of \source\ where differences were detected. 
We identified all cases to be related to the behavior and the setup of~\texttt{TFLiteConverter}. The discrepancy after each repair step is depicted in Figure~\ref{fig:iter-wb-cases}. At each repair step, \tool\ adjusts the weights \& biases of a suspicious layer in \target\ with those from \source, in the order determined by the Layer Activation Analysis (Section~\ref{layer-activations-analysis}), and evaluates  the effects of the change by comparing the predictions of \source\ and \target.

\noindent \textbf{LH:} From the $254$ potential faults \tool\ localized, only $12$ of them were actively affecting the model. These hyperparameters were associated with the \texttt{pads} hyperparameter of \texttt{Conv} and \texttt{MaxPool} nodes (cases \#$11$ \& $13$), as well as 
with the \texttt{axes} of \texttt{Gather} and \texttt{Unsqueeze} nodes responsible for input processing (case \#$14$). Repairing \target\ entailed adjusting their values with those from \source.

\noindent \textbf{CG:}
\tool\ identified $47$ cases of graph-related potential faults. Of these, $4$ cases resulted in fault repairs that improved model performance (reduced output label discrepancy). 
The repairs played a crucial role in conversion cases \#$11$ and \#$13$ that initially presented discrepancies of $100$\% and $35$\%, respectively. Our tool identified that, for case \#$11$, the converter had erroneously introduced multiple output nodes. In addition, three problematic padding nodes were found to affect the behavior of the models.
\tool\ identified such faults and completely repaired the models by (1) removing any additional output nodes added to the model by the converter, and (2) repairing problematic padding nodes.


For all the false positive cases observed in LH and CG, we performed manual inspection and observed that the converters performed semantically equivalent computation graph conversions. 
For instance, in the erroneous conversion presenting the largest number of CG fault localizations --  ResNet101, \texttt{Keras} to \texttt{TFLite} (case \#$8$) -- \source\ contained \texttt{Pad} nodes which were absent in \target. However, \target\ contained a \texttt{pads} hyperparameter in the next convolutional layer that had the same effect. 
Also, \texttt{BatchNorm} nodes of \source\ were replaced by the converter with \texttt{Mul} and \texttt{Add} nodes in \target\, effectively applying the normalization in a two-step semantically equivalent manner. \tool\ attempted to repair these scenarios by replacing them with the corresponding \source\ subgraph but these fixes had no effect on the number of output label discrepancies encountered.

In terms of the number of iterations needed during the repair process, input-based strategies are only applied once. Layer-based strategies typically require multiple iterations.
Four erroneous model conversions that required multiple repair iterations related to WB strategy repair are presented in Figure~\ref{fig:iter-wb-cases}, ranging from $2$ to $7$ repair cycles, using the ILSVRC2017 dataset.

\vspace{-8pt}
\subsection{Fault Localization/Repair Effectiveness}

We consider effectiveness as the number of localized faults that could be repaired to improve \target\ prediction accuracy towards matching this of \source. We go through each fault category to determine the effectiveness of the fault localization component, as seen in the \texttt{Total} row in Table~\ref{tab:faultsdistribution}. 
We find that \tool\ had an $100$\% efficacy rate (all potential fault occurrences detected could be repaired) for preprocessing (PP: $9/9$), input dimensions (ID: $4/4$) and tensor structure (TSS: $2/2$), as well as $98.1$\% efficacy for weights and biases (WB: $430/438$). 
The weights and biases category had the most faults detected and repaired.
\tool\ had much less success on cases related to the computation graph (CG), with $4$ of the $47$ potential faults detected being repaired ($8.5$\%) while other cases when repaired did not result in reduction of output label discrepancy. 
The CG fault repairs, although few in number, were instrumental in completely fixing conversion cases \#$11$ and $13$. The effectiveness for LH category was low ($4$\%), with only $12$ out of the $254$ identified potential faults worth repairing. 
Following manual inspection, we infer that this happens because: (1) graph transformations across \source\ and \target\ may result in syntactic difference but be semantically equivalent. For instance, a \texttt{BatchNorm} node is transformed in \texttt{Mul} and \texttt{Add} nodes. Although this is localized as a potential fault by \tool\, the repair employed by replacing \texttt{Mul} and \texttt{Add} nodes with \texttt{BatchNorm} will not reduce the output label discrepancy and is therefore rejected; (2) the LH difference detected is already mitigated by the converter by introducing additional nodes in the \target. For example, a \texttt{pads} (padding) hyperparameter in a \texttt{Conv} node on \source\ is replaced by a dominating \texttt{Pad} to the correspondent \texttt{Conv} node in target, leading to the same effect.

In terms of the fault analysis for layer-based types, we set \tool\ to perform fault localization with and without fault analysis, and we compared the outcomes. We concluded that \tool\ was able to repair the models in a similar manner with the results presented without the fault analysis, however enabling it applied fewer fault repairs, more effectively avoiding false positive cases where the applied repair was ineffective (up to 3.6\% reduction). We also found the layer ranking assisted towards applying smaller repair steps, with the dissimilarity percentage decreasing more gradually with each repair cycle, making it easy to monitor and control the fault process.


\vspace{-10pt}
\subsection{Repair Validity}

In order to verify our results, we used the ImageNetV2~\cite{imagenetv2} dataset consisting of 10K images which was built with the purpose of providing test data for ImageNet, as an independent validation source. We checked the discrepancies of the repaired \target\ models against \source, and verified that (1) all cases that were fully repaired using ILSVRC2017 also presented no discrepancies with ImageNetV2, and (2) partially repaired cases \#$2$ and $8$ presented $0.28$\% and $0.66$\% discrepancies, lower than those observed in ILSVRC2017. 
The full comparison can be found in Table~\ref{tab:results-repair}. In addition, we used ImageNetV2 to verify that repaired models were giving consistent results against the dataset ground truths. Finally, we tested all $10$ cases shown in Figure~\ref{fig:conversionsorig} presenting no discrepancies with ImageNetV2, to verify that no additional errors were detected.
In detail, we verified in our experiments that the results of each repair cycle do not misclassify previously correct results, by testing all repaired models on ImageNetV2, checking the classification in two ways - against ground truth for the dataset inputs, and by comparing the results between source and target models. We did not observe any misclassification cases of correct \target\ model predictions prior to the repair process across the entirety of our experiment set.

\vspace{-7pt}
\subsection{Root Causes of Faults}
We consider that many of the errors detected might have common root causes as part of the conversion process. Upon further manual examination, we observed that (1) all faults related with preprocessing could be repaired when the proper setting was selected, (2) errors related to Weights \& Biases (WB) were related to an implicit action caused by \texttt{tf2onnx} converter, and (3) issues related to Input Dimensions (ID), Tensor Shape \& Structure (TSS), Layer Hyperparameters (LH) and Computational Graph (CG) were attributed to problematic handling of the layers by the conversion tools (e.g., both \texttt{torch.onnx} and  \texttt{onnx2keras} converters  mishandled model inputs). We aim to report such issues to the development teams of each converter tool.

\vspace{-8pt}
\subsection{Results Generalizability}
We expect our approach to be applicable to any DL frameworks containing core computational layers and their conversion tools - including non-CNN models, requiring from the user to specify the core computational layers to be used in the comparison. In addition, the user must define a comparator for the evaluation of the model type under test (e.g., BLEU~\cite{BLEU} for text generation models). Providing proof of \tool\ effectiveness in this scenario, is subject to future work. In terms of our experiments, we chose image recognition models in our experiments as they are more tractable and need less resources than other DL architectures like transformers, as our experiments involved $15$ model conversion cases with numerous iterations and analysis within each. All the models selected are widely used, either as-is or as the base for extension of models deployed on more complex tasks (e.g., ResNet101 extended for object detection). Our current implementation is bassed on ONNX, as it is a widely-used format, providing a comprehensive API for model definition and modification. 
 However, our methodology can be easily extended to other representations.

In terms of the fault types supported by \tool, they are based on a survey of issues encountered in discussion forums and empirical investigations (Table~\ref{tab:faults-class}). There may be other fault types not currently considered by \tool. We are aware of this and we expect that \tool\ will keep evolving to handle emerging issues and other fault categories in model conversions between DL frameworks. 

\begin{table}[t]
    \centering
    \caption{Experiments Overview: Discrepancies between \source\ and \target. before and after the repair process. for ILSVRC2017 and ImageNetV2 datasets.}
    \vspace{-7pt}
    \label{tab:results-repair}
    \small
    \begin{tabular}{|l|l|l|l|l|l|}
    \hline
        ~ & ~ & \multicolumn{2}{|c|}{\textbf{ILSVRC2017}} & \multicolumn{2}{|c|}{\textbf{ImageNetV2}}\\
        \hline
        \# & \textbf{MobileNetV2} & \textbf{Initial} & \textbf{Fixed} & \textbf{Initial} & \textbf{Fixed} \\ \hline
        1 & Keras-to-Torch & 83.41\% & \textbf{0\%} & 74.85\% & \textbf{0\%} \\ \hline
        2 & Keras-to-TFLite & 49.34\% & \textbf{0.33\%} & 31.87\% & \textbf{0.28\%} \\ \hline
        3 & Torch-to-TF & 68.21\% & \textbf{0\%} & 53.77\% & \textbf{0\%} \\ \hline
        4 & Torch-to-Keras & 47.21\% & \textbf{0\%} & 26.44\% & \textbf{0\%} \\ \hline
        5 & TF-to-TFLite & 9.70\% & \textbf{0\%} & 6.32\% & \textbf{0\%} \\ \hline
        6 & TF-to-Keras & 48.90\% & \textbf{0\%} & 31.56\% & \textbf{0\%} \\ \hline
        ~ & ~ & ~ & ~ & ~ & ~ \\ \hline
        ~ & \textbf{ResNet101} & ~ & ~ & ~ & ~ \\ \hline
        7 & Keras-to-Torch & 36\% & \textbf{0\%} & 89\% & \textbf{0\%} \\ \hline
        8 & Keras-to-TFLite & 3.12\% & \textbf{1.50\%} & 1.58\% & \textbf{0.66\%} \\ \hline
        9 & Torch-to-TFLite & 72.30\% & \textbf{0\%} & 53.41\% & \textbf{0\%} \\ \hline
        10 & Torch-to-TF & 73.90\% & \textbf{0\%} & 56.15\% & \textbf{0\%} \\ \hline
        11 & TFLite-to-Keras & 100.00\% & \textbf{0\%} & 100.00\% & \textbf{0\%} \\ \hline
        12 & TF-to-TFLite & 2.70\% & 2.70\% & 1.32\% & 1.32\% \\ \hline
        13 & TF-to-Keras & 34.85\% & \textbf{0\%} & 17.59\% & \textbf{0\%} \\ \hline
        ~ & ~ & ~ & ~ & ~ & ~ \\ \hline
        ~ & \textbf{InceptionV3} & ~ & ~ & ~ & ~ \\ \hline
        14 & Torch-to-Keras & 32.38\% & \textbf{0\%} & 15.89\% & \textbf{0\%} \\ \hline
        15 & TF-to-TFLite & 4.36\% & \textbf{0\%} & 1.53\% & \textbf{0\%} \\ \hline
    \end{tabular}
    \vspace{3pt}
\end{table}

\vspace{-5pt}
\section{Challenges \& Limitations}
\label{challenges}
Upon \tool\ development, we encountered a number of challenges and limitation, which we describe in this section.
\\
\noindent \textbf{Node  Matching Across Source \& Target} In order to perform the fault localization and repair process, \tool\ had to correctly match the nodes across \source\ and \target. However, the node matching was not one-to-one, due to (1) changes in graph structure applied by converter tools, (2) modifications in shape and internal structure of model nodes and (3) alterations in layer names between the model variants. 
To mitigate this problem, we introduced a solid set of requirements that had to be fulfilled for two nodes to be marked as the same. In detail, we required that two model nodes were (1) of the same type, (2) of the same shape, and (3) they shared a number of parameters on which values could only deviate by a small margin of difference (absolute value of $0.2$, obtained following step by step decrease to the point where the system failed to match the same nodes). 
For the moment however, we match only core computational layers (e.g., convolutions), considering that their number is guaranteed to remain unaffected by the conversion process due to model design. We also limit the matching algorithm to nodes containing weights and/or biases, as the presence of such properties increased significantly the matching ability of \tool, to the extent of more than 95\% across \source\ and \target. We aim to improve our system to this direction by performing matching considering the presence of the same hyperparameters across nodes between \source\ and \target.
\\
\noindent \textbf{ONNX Opset Representation Limitations:} For case \#$12$, we were unable to compile both \source\ and \target\ to the same opset, despite our efforts, due to ONNX operation incompatibilities. For that reason, we generated \source\ in \texttt{opset=11} and \target\ in \texttt{opset=13}. However, the two opsets had completely different graph representations, and \tool\ was unable to correctly match the layers across model variants and perform fault localization and repair.

\noindent \textbf{Computationally Intensive Fault Analysis:}
As the fault analysis we conducted performed per-element comparison for a number of inputs across each layer, it resulted in a vast amount of computations. This limited our tool from performing it to the full validation dataset, for reasons of tractability. For that matter, we selected a random sample of up to 100 images and we conducted our analysis on it, and repeated this process 3 times to solidify our results. We also accelerated the process by enabling parallel computation for the analysis, using \texttt{Joblib} python library~\cite{joblib}.

\vspace{-5pt}
\section{Conclusion}

We presented \tool, an automated approach for fault localization and repair during model conversion between deep learning frameworks. 
\tool\ is capable of detecting and fixing six commonly encountered fault types that are input and layer-based related to pre-processing, input dimensions, layer weights, biases, hyperparameters, and computation graph. 
We evaluated the effectiveness of \tool\ in fixing previously reported model conversion bugs~\cite{louloudakis2023deltann} for three widely used image recognition models converted across four different deep learning frameworks. 
We found that \tool\ was highly effective in repairing the erroneous model conversions reported by completely fixing or significantly improving $14$ out of the $15$ cases under test, repairing a total of $462$ bugs.

\vspace{-5pt}
\begin{acks}
The research has been funded by the Royal Society Industry Fellowship "AutoTest: Testing Autonomous Vehicle Perception Safety on Hardware Accelerators" and Huawei Edinburgh Joint Lab Project "RobustCheck: Testing Robustness of Compiler Optimisations and Deep Learning Frameworks".
\end{acks}

\eject \vfill
\balance

\bibliographystyle{ACM-Reference-Format}
\bibliography{0-main}


\begin{thebibliography}{80}


\ifx \showCODEN    \undefined \def \showCODEN     #1{\unskip}     \fi
\ifx \showDOI      \undefined \def \showDOI       #1{#1}\fi
\ifx \showISBNx    \undefined \def \showISBNx     #1{\unskip}     \fi
\ifx \showISBNxiii \undefined \def \showISBNxiii  #1{\unskip}     \fi
\ifx \showISSN     \undefined \def \showISSN      #1{\unskip}     \fi
\ifx \showLCCN     \undefined \def \showLCCN      #1{\unskip}     \fi
\ifx \shownote     \undefined \def \shownote      #1{#1}          \fi
\ifx \showarticletitle \undefined \def \showarticletitle #1{#1}   \fi
\ifx \showURL      \undefined \def \showURL       {\relax}        \fi
\providecommand\bibfield[2]{#2}
\providecommand\bibinfo[2]{#2}
\providecommand\natexlab[1]{#1}
\providecommand\showeprint[2][]{arXiv:#2}

\bibitem[\protect\citeauthoryear{??}{tfl}{2020}]%
        {tflite2onnx}
 \bibinfo{year}{2020}\natexlab{}.
\newblock \bibinfo{title}{{tflite2onnx}}.
\newblock \bibinfo{howpublished}{https://github.com/zhenhuaw-me/tflite2onnx}.
\newblock
\urldef\tempurl%
\url{https://github.com/zhenhuaw-me/tflite2onnx}
\showURL{%
\tempurl}
\newblock
\shownote{[Accessed 6-June-2023].}


\bibitem[\protect\citeauthoryear{??}{TF-}{2021}]%
        {TF-errors}
 \bibinfo{year}{2021}\natexlab{}.
\newblock \bibinfo{title}{{TensorFlow - Frequently Asked Questions}}.
\newblock
\newblock
\urldef\tempurl%
\url{https://www.tensorflow.org/lite/guide/faq}
\showURL{%
\tempurl}


\bibitem[\protect\citeauthoryear{??}{tf2}{2022}]%
        {tf2onnx}
 \bibinfo{year}{2022}\natexlab{}.
\newblock \bibinfo{title}{{TF2ONNX}}.
\newblock \bibinfo{howpublished}{https://github.com/onnx/tensorflow-onnx}.
\newblock
\newblock
\shownote{[Accessed 15-Feb-2023].}


\bibitem[\protect\citeauthoryear{??}{sta}{2023a}]%
        {stackoverflow_accuracy_drop_tensorflow_to_tflite_2023}
 \bibinfo{year}{2023}\natexlab{a}.
\newblock \bibinfo{title}{Accuracy Drop Between {TensorFlow} Model and Converted {TFLite}}.
\newblock \bibinfo{howpublished}{https://stackoverflow.com/questions/65731362/accuracy-drop-between-tensorflow-model-and-converted-tflite}.
\newblock
\newblock
\shownote{Accessed 13 Dec. 2023.}


\bibitem[\protect\citeauthoryear{??}{ten}{2023a}]%
        {tensorflow_forum_accuracy_drop_2023}
 \bibinfo{year}{2023}\natexlab{a}.
\newblock \bibinfo{title}{After Converting the Model to .tflite and Running It on {Android}, the Accuracy Drops}.
\newblock \bibinfo{howpublished}{https://discuss.tensorflow.org/t/after-converting-the-model-to-tflite-and-running-it-on-android-the-accuracy-drops/1310/2}.
\newblock
\newblock
\shownote{Accessed 13 Dec. 2023.}


\bibitem[\protect\citeauthoryear{??}{sta}{2023b}]%
        {stackoverflow_tflite_2023}
 \bibinfo{year}{2023}\natexlab{b}.
\newblock \bibinfo{title}{Always Getting 0 for Prediction from Tensorflow Lite Model}.
\newblock \bibinfo{howpublished}{\url{https://stackoverflow.com/questions/67069167/always-getting-0-for-prediction-from-tensorflow-lite-model}}.
\newblock
\newblock
\shownote{Accessed 13 Dec. 2023.}


\bibitem[\protect\citeauthoryear{??}{git}{2023a}]%
        {github_pytorch2keras_issue135_2023}
 \bibinfo{year}{2023}\natexlab{a}.
\newblock \bibinfo{title}{Batch Normalization Layers Not Present in the Converted Keras Model - Issue \#135 - Gmalivenko/pytorch2keras}.
\newblock \bibinfo{howpublished}{\url{https://github.com/gmalivenko/pytorch2keras/issues/135}}.
\newblock
\newblock
\shownote{Accessed 13 Dec. 2023.}


\bibitem[\protect\citeauthoryear{??}{sta}{2023c}]%
        {stackoverflow_dimension_mismatch_2023}
 \bibinfo{year}{2023}\natexlab{c}.
\newblock \bibinfo{title}{Cannot Set Tensor: Dimension Mismatch}.
\newblock \bibinfo{howpublished}{\url{https://stackoverflow.com/questions/72516622/cannot-set-tensor-dimension-mismatch}}.
\newblock
\newblock
\shownote{Accessed 13 Dec. 2023.}


\bibitem[\protect\citeauthoryear{??}{pyt}{2023}]%
        {pytorch_forum_coreml_wrong_prediction_2020}
 \bibinfo{year}{2023}\natexlab{}.
\newblock \bibinfo{title}{Convert to CoreML but Predict Wrong}.
\newblock \bibinfo{howpublished}{\url{https://discuss.pytorch.org/t/convert-to-coreml-but-predict-wrong/66355/3}}.
\newblock
\newblock
\shownote{Accessed 13 Dec. 2023.}


\bibitem[\protect\citeauthoryear{??}{git}{2023b}]%
        {github_pytorch2keras_issue127_2023}
 \bibinfo{year}{2023}\natexlab{b}.
\newblock \bibinfo{title}{Converted Keras Model Has Different Parameters - Issue \#127 - Gmalivenko/pytorch2keras}.
\newblock \bibinfo{howpublished}{\url{https://github.com/gmalivenko/pytorch2keras/issues/127}}.
\newblock
\newblock
\shownote{Accessed 13 Dec. 2023.}


\bibitem[\protect\citeauthoryear{??}{git}{2023c}]%
        {github_pytorch2keras_issue124_2023}
 \bibinfo{year}{2023}\natexlab{c}.
\newblock \bibinfo{title}{Converted Model Has Different Weights Than the Original Model - Issue \#124 - Gmalivenko/pytorch2keras}.
\newblock \bibinfo{howpublished}{\url{https://github.com/gmalivenko/pytorch2keras/issues/124}}.
\newblock
\newblock
\shownote{Accessed 13 Dec. 2023.}


\bibitem[\protect\citeauthoryear{??}{sta}{2023d}]%
        {stackoverflow_coreml_wrong_prediction_2023a}
 \bibinfo{year}{2023}\natexlab{d}.
\newblock \bibinfo{title}{{CoreML} Model Converted from {PyTorch} Model Giving the Wrong Prediction Probability}.
\newblock \bibinfo{howpublished}{https://stackoverflow.com/questions/64519191/coreml-model-converted-from-pytorch-model-giving-the-wrong-prediction-probabilit}.
\newblock
\newblock
\shownote{Accessed 13 Dec. 2023.}


\bibitem[\protect\citeauthoryear{??}{git}{2023d}]%
        {github_onnx_keras_issue572_2023}
 \bibinfo{year}{2023}\natexlab{d}.
\newblock \bibinfo{title}{Custom Converter Being Wrapped by Transpose Statements (set\_converter) - Issue \#572 - Onnx/keras-onnx}.
\newblock \bibinfo{howpublished}{\url{https://github.com/onnx/keras-onnx/issues/572}}.
\newblock
\newblock
\shownote{Accessed 13 Dec. 2023.}


\bibitem[\protect\citeauthoryear{??}{git}{2023e}]%
        {github_microsoft_MMdnn_issue823_2023}
 \bibinfo{year}{2023}\natexlab{e}.
\newblock \bibinfo{title}{Different Accuracy After Model Conversion from {Keras} to {Caffe} - Issue \#823 - {Microsoft/MMdnn}}.
\newblock \bibinfo{howpublished}{\url{https://github.com/microsoft/MMdnn/issues/823}}.
\newblock
\newblock
\shownote{Accessed 13 Dec. 2023.}


\bibitem[\protect\citeauthoryear{??}{sta}{2023e}]%
        {stackoverflow_dimension_mismatch_keras_to_onnx_2023}
 \bibinfo{year}{2023}\natexlab{e}.
\newblock \bibinfo{title}{Dimension Mismatch during {Keras} to {ONNX} Conversion ({2D} Output)}.
\newblock \bibinfo{howpublished}{\url{https://stackoverflow.com/questions/70861809/dimension-mismatch-during-keras-to-onnx-conversion-2d-output}}.
\newblock
\newblock
\shownote{Accessed 13 Dec. 2023.}


\bibitem[\protect\citeauthoryear{??}{git}{2023f}]%
        {github_pytorch2keras_issue31_2023}
 \bibinfo{year}{2023}\natexlab{f}.
\newblock \bibinfo{title}{Error: Failing in Transpose Layer (Cannot Permute Batch Dimension. Result May Be Wrong) - Issue \#31 - Gmalivenko/pytorch2keras}.
\newblock \bibinfo{howpublished}{\url{https://github.com/gmalivenko/pytorch2keras/issues/31}}.
\newblock
\newblock
\shownote{Accessed 13 Dec. 2023.}


\bibitem[\protect\citeauthoryear{??}{ten}{2023b}]%
        {tensorflow_forum_model_accuracy_loss_2023}
 \bibinfo{year}{2023}\natexlab{b}.
\newblock \bibinfo{title}{Extreme Model Accuracy Loss Due to {TFLite} Conversion W/ Quantization}.
\newblock \bibinfo{howpublished}{\url{https://discuss.tensorflow.org/t/extreme-model-accuracy-loss-due-to-tflite-conversion-w-quantization/2637/5}}.
\newblock
\newblock
\shownote{Accessed 13 Dec. 2023.}


\bibitem[\protect\citeauthoryear{??}{git}{2023g}]%
        {github_tensorflow_issue35194_2023}
 \bibinfo{year}{2023}\natexlab{g}.
\newblock \bibinfo{title}{Failed to Convert Weights to 8 Bit Precision: "Quantize Weights Tool Only Supports Tflite Models with One Subgraph" - Issue \#35194 - Tensorflow/tensorflow}.
\newblock \bibinfo{howpublished}{\url{https://github.com/tensorflow/tensorflow/issues/35194}}.
\newblock
\newblock
\shownote{Accessed 13 Dec. 2023.}


\bibitem[\protect\citeauthoryear{??}{git}{2023h}]%
        {github_onnx2keras_issue135_2023}
 \bibinfo{year}{2023}\natexlab{h}.
\newblock \bibinfo{title}{How to Modify the Convolution Property to Same. - Issue \#135 - Gmalivenko/onnx2keras}.
\newblock \bibinfo{howpublished}{\url{https://github.com/gmalivenko/onnx2keras/issues/135}}.
\newblock
\newblock
\shownote{Accessed 13 Dec. 2023.}


\bibitem[\protect\citeauthoryear{??}{git}{2023i}]%
        {github_sagemaker_python_sdk_issue613_2023}
 \bibinfo{year}{2023}\natexlab{i}.
\newblock \bibinfo{title}{Hyperparameter Values Forcefully Converted to Strings, Thus Unable to Pass a List - Issue \#613 - Aws/sagemaker-python-sdk}.
\newblock \bibinfo{howpublished}{\url{https://github.com/aws/sagemaker-python-sdk/issues/613}}.
\newblock
\newblock
\shownote{Accessed 13 Dec. 2023.}


\bibitem[\protect\citeauthoryear{??}{sta}{2023f}]%
        {stackoverflow_tf_incorrect_data_2023}
 \bibinfo{year}{2023}\natexlab{f}.
\newblock \bibinfo{title}{Incorrect Data Response in Tensorflow}.
\newblock \bibinfo{howpublished}{\url{https://stackoverflow.com/questions/76418614/incorrect-data-response-in-tensorflow}}.
\newblock
\newblock
\shownote{Accessed 13 Dec. 2023.}


\bibitem[\protect\citeauthoryear{??}{git}{2023j}]%
        {github_pytorch2keras_dimension_issue_2023}
 \bibinfo{year}{2023}\natexlab{j}.
\newblock \bibinfo{title}{Keras Model's Summary, First Output Shape is [(None, 1, 28, 28)] - Issue \#104 - Gmalivenko/pytorch2keras}.
\newblock \bibinfo{howpublished}{\url{https://github.com/gmalivenko/pytorch2keras/issues/104}}.
\newblock
\newblock
\shownote{Accessed 13 Dec. 2023.}


\bibitem[\protect\citeauthoryear{??}{git}{2023k}]%
        {github_pytorch2keras_issue78_2023}
 \bibinfo{year}{2023}\natexlab{k}.
\newblock \bibinfo{title}{Layer Weight Shape Don't Match - Issue \#78 - Gmalivenko/pytorch2keras}.
\newblock \bibinfo{howpublished}{\url{https://github.com/gmalivenko/pytorch2keras/issues/78}}.
\newblock
\newblock
\shownote{Accessed 13 Dec. 2023.}


\bibitem[\protect\citeauthoryear{??}{git}{2023l}]%
        {github_tensorflow_onnx_issue1203_2023}
 \bibinfo{year}{2023}\natexlab{l}.
\newblock \bibinfo{title}{Missing Shape Information for 'NonZero' Node Derived from 'Where' Node - Issue \#1203 - Onnx/tensorflow-onnx}.
\newblock \bibinfo{howpublished}{\url{https://github.com/onnx/tensorflow-onnx/issues/1203}}.
\newblock
\newblock
\shownote{Accessed 13 Dec. 2023.}


\bibitem[\protect\citeauthoryear{??}{sta}{2023g}]%
        {stackoverflow_model_dimension_error_2023}
 \bibinfo{year}{2023}\natexlab{g}.
\newblock \bibinfo{title}{Model Gets Correct Input Dimensions, but Throws Dimension Error}.
\newblock \bibinfo{howpublished}{\url{https://stackoverflow.com/questions/56292213/model-gets-correct-input-dimensions-but-throws-dimension-error}}.
\newblock
\newblock
\shownote{Accessed 13 Dec. 2023.}


\bibitem[\protect\citeauthoryear{??}{onn}{2023a}]%
        {onnx2keras}
 \bibinfo{year}{2023}\natexlab{a}.
\newblock \bibinfo{title}{{onnx2keras}}.
\newblock \bibinfo{howpublished}{https://github.com/gmalivenko/onnx2keras}.
\newblock
\newblock
\shownote{[Accessed 15-Feb-2023].}


\bibitem[\protect\citeauthoryear{??}{onn}{2023b}]%
        {onnx2torch}
 \bibinfo{year}{2023}\natexlab{b}.
\newblock \bibinfo{title}{{onnx2torch}}.
\newblock \bibinfo{howpublished}{https://github.com/ENOT-AutoDL/onnx2torch}.
\newblock
\newblock
\shownote{[Accessed 15-Feb-2023].}


\bibitem[\protect\citeauthoryear{??}{onn}{2023c}]%
        {onnxsite}
 \bibinfo{year}{2023}\natexlab{c}.
\newblock \bibinfo{title}{{Open Neural Network Exchange}}.
\newblock \bibinfo{howpublished}{https://onnx.ai/}.
\newblock
\newblock
\shownote{[Accessed 8-Dec-2023].}


\bibitem[\protect\citeauthoryear{??}{onn}{2023d}]%
        {onnxmltools}
 \bibinfo{year}{2023}\natexlab{d}.
\newblock \bibinfo{title}{{Open Neural Network Exchange Tools - GitHub}}.
\newblock \bibinfo{howpublished}{https://github.com/onnx/onnxmltools}.
\newblock
\newblock
\shownote{[Accessed 8-Dec-2023].}


\bibitem[\protect\citeauthoryear{??}{sta}{2023h}]%
        {stackoverflow_poor_tflite_accuracy_android_2023}
 \bibinfo{year}{2023}\natexlab{h}.
\newblock \bibinfo{title}{Poor {TensorFlow Lite} Accuracy in {Android} Application}.
\newblock \bibinfo{howpublished}{https://stackoverflow.com/questions/69352192/poor-tensorflow-lite-accuracy-in-android-application}.
\newblock
\newblock
\shownote{Accessed 13 Dec. 2023.}


\bibitem[\protect\citeauthoryear{??}{git}{2023m}]%
        {github_onnx_tensorflow_conversion_issue_2023}
 \bibinfo{year}{2023}\natexlab{m}.
\newblock \bibinfo{title}{PyTorch to ONNX to TensorFlow - How to Convert from NCHW (ONNX) to NHWC (TensorFlow Lite) - Issue \#862 - Onnx/onnx-tensorflow}.
\newblock \bibinfo{howpublished}{\url{https://github.com/onnx/onnx-tensorflow/issues/862}}.
\newblock
\newblock
\shownote{Accessed 13 Dec. 2023.}


\bibitem[\protect\citeauthoryear{??}{git}{2023n}]%
        {github_pytorch2keras_issue76_2023}
 \bibinfo{year}{2023}\natexlab{n}.
\newblock \bibinfo{title}{Reshape After View is Wrong - Issue \#76 - Gmalivenko/pytorch2keras}.
\newblock \bibinfo{howpublished}{\url{https://github.com/gmalivenko/pytorch2keras/issues/76}}.
\newblock
\newblock
\shownote{Accessed 13 Dec. 2023.}


\bibitem[\protect\citeauthoryear{??}{sta}{2023i}]%
        {stackoverflow_strange_dimension_behaviour_2023}
 \bibinfo{year}{2023}\natexlab{i}.
\newblock \bibinfo{title}{Strange Dimension Behaviour: Needs Both Dimension 2 and 3, Unsure Why}.
\newblock \bibinfo{howpublished}{https://stackoverflow.com/questions/58031343/strange-dimension-behaviour-needs-both-dimension-2-and-3-unsure-why}.
\newblock
\newblock
\shownote{Accessed 13 Dec. 2023.}


\bibitem[\protect\citeauthoryear{??}{sta}{2023j}]%
        {stackoverflow_tflite_conversion_changes_weights_2023a}
 \bibinfo{year}{2023}\natexlab{j}.
\newblock \bibinfo{title}{{TensorFlow Lite} Conversion Changes Model Weights}.
\newblock \bibinfo{howpublished}{https://stackoverflow.com/questions/54404262/tensorflow-lite-conversion-changes-model-weights}.
\newblock
\newblock
\shownote{Accessed 13 Dec. 2023.}


\bibitem[\protect\citeauthoryear{??}{git}{2023o}]%
        {github_microsoft_MMdnn_issue831_2023}
 \bibinfo{year}{2023}\natexlab{o}.
\newblock \bibinfo{title}{{Tensorflow} to {Caffe}, Reshape Layer - Issue \#831 - {Microsoft/MMdnn}}.
\newblock \bibinfo{howpublished}{\url{https://github.com/microsoft/MMdnn/issues/831}}.
\newblock
\newblock
\shownote{Accessed 13 Dec. 2023.}


\bibitem[\protect\citeauthoryear{??}{sta}{2023k}]%
        {stackoverflow_valueerror_dimension_mismatch_2_2023}
 \bibinfo{year}{2023}\natexlab{k}.
\newblock \bibinfo{title}{{TensorFlow}/{Keras} ValueError on Input Shape}.
\newblock \bibinfo{howpublished}{\url{https://stackoverflow.com/questions/68837658/tensorflow-keras-valueerror-on-input-shape}}.
\newblock
\newblock
\shownote{Accessed 13 Dec. 2023.}


\bibitem[\protect\citeauthoryear{??}{git}{2023p}]%
        {github_tensorflow_issue31205_2023}
 \bibinfo{year}{2023}\natexlab{p}.
\newblock \bibinfo{title}{{TFLite}: Changing Weights - Issue \#31205 - Tensorflow/tensorflow}.
\newblock \bibinfo{howpublished}{\url{https://github.com/tensorflow/tensorflow/issues/31205}}.
\newblock
\newblock
\shownote{Accessed 13 Dec. 2023.}


\bibitem[\protect\citeauthoryear{??}{sta}{2023l}]%
        {stackoverflow_tflite_model_overflows_on_gpu_2023}
 \bibinfo{year}{2023}\natexlab{l}.
\newblock \bibinfo{title}{{TFLite Model Overflows on {GPU}, OK on {CPU} - What Are the Differences Internally?}}
\newblock \bibinfo{howpublished}{https://stackoverflow.com/questions/62032560/tflite-model-overflows-on-gpu-ok-on-cpu-what-are-the-differences-internally}.
\newblock
\newblock
\shownote{Accessed 13 Dec. 2023.}


\bibitem[\protect\citeauthoryear{??}{git}{2023q}]%
        {github_tensorflow_issue31359_2023}
 \bibinfo{year}{2023}\natexlab{q}.
\newblock \bibinfo{title}{{TFLite} Output Different Result with Pbfile when Using Only One Convolutional Layer? - Issue \#31359 - Tensorflow/tensorflow}.
\newblock \bibinfo{howpublished}{\url{https://github.com/tensorflow/tensorflow/issues/31359}}.
\newblock
\newblock
\shownote{Accessed 13 Dec. 2023.}


\bibitem[\protect\citeauthoryear{??}{tvm}{2023}]%
        {tvm_debugger}
 \bibinfo{year}{2023}\natexlab{}.
\newblock \bibinfo{title}{{TVM Debugger}}.
\newblock \bibinfo{howpublished}{https://tvm.apache.org/docs/arch/debugger.html}.
\newblock
\newblock
\shownote{[Accessed 13-Dec-2023].}


\bibitem[\protect\citeauthoryear{??}{ten}{2023c}]%
        {tensorflow_forum_dimension_mismatch_2023}
 \bibinfo{year}{2023}\natexlab{c}.
\newblock \bibinfo{title}{{ValueError}: Cannot Set Tensor: Dimension Mismatch}.
\newblock \bibinfo{howpublished}{https://discuss.tensorflow.org/t/valueerror-cannot-set-tensor-dimension-mismatch/15313}.
\newblock
\newblock
\shownote{Accessed 13 Dec. 2023.}


\bibitem[\protect\citeauthoryear{??}{sta}{2023m}]%
        {stackoverflow_valueerror_dimension_mismatch_1_2023}
 \bibinfo{year}{2023}\natexlab{m}.
\newblock \bibinfo{title}{{ValueError}: Cannot Set Tensor: Dimension Mismatch (3 but Expected 4)}.
\newblock \bibinfo{howpublished}{https://stackoverflow.com/questions/67068742/valueerror-cannot-set-tensor-dimension-mismatch-got-3-but-expected-4-for-inpu}.
\newblock
\newblock
\shownote{Accessed 13 Dec. 2023.}


\bibitem[\protect\citeauthoryear{??}{git}{2023r}]%
        {github_onnx_tensorflow_issue2246_2023}
 \bibinfo{year}{2023}\natexlab{r}.
\newblock \bibinfo{title}{{ValueError}: Graph Has Cycles - Issue \#2246 - Onnx/tensorflow-onnx}.
\newblock \bibinfo{howpublished}{\url{https://github.com/onnx/tensorflow-onnx/issues/2246}}.
\newblock
\newblock
\shownote{Accessed 13 Dec. 2023.}


\bibitem[\protect\citeauthoryear{??}{sta}{2023n}]%
        {stackoverflow_tf_problem_or_wrong_2023}
 \bibinfo{year}{2023}\natexlab{n}.
\newblock \bibinfo{title}{Want to Confirm if This Is a Problem with Model or I Am Doing Something Wrong {(TF)}}.
\newblock \bibinfo{howpublished}{https://stackoverflow.com/questions/73431543/want-to-confirm-if-this-is-a-problem-with-model-or-i-am-doing-something-wrong-tf}.
\newblock
\newblock
\shownote{Accessed 13 Dec. 2023.}


\bibitem[\protect\citeauthoryear{??}{git}{2023s}]%
        {github_microsoft_MMdnn_issue297_2023}
 \bibinfo{year}{2023}\natexlab{s}.
\newblock \bibinfo{title}{Weights Are Not Equal when Convert Model from {Tensorflow} to {Caffe} - Issue \#297 - {Microsoft/MMdnn}}.
\newblock \bibinfo{howpublished}{\url{https://github.com/microsoft/MMdnn/issues/297}}.
\newblock
\newblock
\shownote{Accessed 13 Dec. 2023.}


\bibitem[\protect\citeauthoryear{??}{git}{2023t}]%
        {github_onnx_tensorflow_issue782_2023}
 \bibinfo{year}{2023}\natexlab{t}.
\newblock \bibinfo{title}{Why Does Onnx-tensorflow Add Transpose Layers for Each Conv2D Layer? - Issue \#782 - Onnx/onnx-tensorflow}.
\newblock \bibinfo{howpublished}{\url{https://github.com/onnx/onnx-tensorflow/issues/782}}.
\newblock
\newblock
\shownote{Accessed 13 Dec. 2023.}


\bibitem[\protect\citeauthoryear{??}{man}{2024a}]%
        {manualconversion2}
 \bibinfo{year}{2024}\natexlab{a}.
\newblock \bibinfo{title}{Convert {Keras} model to {PyTorch}}.
\newblock \bibinfo{howpublished}{\url{https://stackoverflow.com/questions/68413480/convert-keras-model-to-pytorch}}.
\newblock
\newblock
\shownote{Accessed 23 Feb. 2024.}


\bibitem[\protect\citeauthoryear{??}{job}{2024}]%
        {joblib}
 \bibinfo{year}{2024}\natexlab{}.
\newblock \bibinfo{title}{{Joblib}}.
\newblock \bibinfo{howpublished}{https://joblib.readthedocs.io/en/latest/parallel.html}.
\newblock
\newblock
\shownote{[Accessed 9-Jul-2024].}


\bibitem[\protect\citeauthoryear{??}{man}{2024b}]%
        {manualconversion1}
 \bibinfo{year}{2024}\natexlab{b}.
\newblock \bibinfo{title}{{PyTorch} to {Keras} code equivalence}.
\newblock \bibinfo{howpublished}{\url{https://stackoverflow.com/questions/46866763/pytorch-to-keras-code-equivalence}}.
\newblock
\newblock
\shownote{Accessed 23 Feb. 2024.}


\bibitem[\protect\citeauthoryear{??}{ten}{2024}]%
        {tensorflow-lite-model-gives-very-different-accuracy-value-compared-to-python-mod}
 \bibinfo{year}{2024}\natexlab{}.
\newblock \bibinfo{title}{{TensorFlow Lite} model gives very different accuracy value compared to python model}.
\newblock \bibinfo{howpublished}{https://stackoverflow.com/questions/52057552/tensorflow-lite-model-gives-very-different-accuracy-value-compared-to-python-mod}.
\newblock
\newblock
\shownote{Accessed 23 Mar. 2024.}


\bibitem[\protect\citeauthoryear{ACervantes}{ACervantes}{2021}]%
        {nvidia_forum_strides_problem_2021}
\bibfield{author}{\bibinfo{person}{ACervantes}.} \bibinfo{year}{2021}\natexlab{}.
\newblock \bibinfo{title}{Strides Problem on the Nvconverter}.
\newblock
\newblock
\urldef\tempurl%
\url{https://forums.developer.nvidia.com/t/strides-problem-on-the-nvconverter/53500}
\showURL{%
\tempurl}
\newblock
\shownote{Accessed 13 Dec. 2023.}


\bibitem[\protect\citeauthoryear{Chen, Moreau, Jiang, Zheng, Yan, Shen, Cowan, Wang, Hu, Ceze, Guestrin, and Krishnamurthy}{Chen et~al\mbox{.}}{2018}]%
        {tvm}
\bibfield{author}{\bibinfo{person}{Tianqi Chen}, \bibinfo{person}{Thierry Moreau}, \bibinfo{person}{Ziheng Jiang}, \bibinfo{person}{Lianmin Zheng}, \bibinfo{person}{Eddie Yan}, \bibinfo{person}{Haichen Shen}, \bibinfo{person}{Meghan Cowan}, \bibinfo{person}{Leyuan Wang}, \bibinfo{person}{Yuwei Hu}, \bibinfo{person}{Luis Ceze}, \bibinfo{person}{Carlos Guestrin}, {and} \bibinfo{person}{Arvind Krishnamurthy}.} \bibinfo{year}{2018}\natexlab{}.
\newblock \showarticletitle{{{{TVM}}: {{An}} Automated End-to-End Optimizing Compiler for Deep Learning}}. In \bibinfo{booktitle}{\emph{13th {{USENIX}} Symposium on Operating Systems Design and Implementation ({{OSDI}} 18)}}. \bibinfo{pages}{578--594}.
\newblock
\showISBNx{978-1-939133-08-3}


\bibitem[\protect\citeauthoryear{Chen, Cao, Liu, Wang, Xie, and Liu}{Chen et~al\mbox{.}}{2020}]%
        {chen2020comprehensive}
\bibfield{author}{\bibinfo{person}{Zhenpeng Chen}, \bibinfo{person}{Yanbin Cao}, \bibinfo{person}{Yuanqiang Liu}, \bibinfo{person}{Haoyu Wang}, \bibinfo{person}{Tao Xie}, {and} \bibinfo{person}{Xuanzhe Liu}.} \bibinfo{year}{2020}\natexlab{}.
\newblock \showarticletitle{{A Comprehensive Study on Challenges in Deploying Deep Learning Based Software}}. In \bibinfo{booktitle}{\emph{Proceedings of the 28th ACM Joint Meeting on European Software Engineering Conference and Symposium on the Foundations of Software Engineering}}. \bibinfo{pages}{750--762}.
\newblock


\bibitem[\protect\citeauthoryear{Chen, Yao, Lou, Cao, Liu, Wang, and Liu}{Chen et~al\mbox{.}}{2021}]%
        {chen2021empirical}
\bibfield{author}{\bibinfo{person}{Zhenpeng Chen}, \bibinfo{person}{Huihan Yao}, \bibinfo{person}{Yiling Lou}, \bibinfo{person}{Yanbin Cao}, \bibinfo{person}{Yuanqiang Liu}, \bibinfo{person}{Haoyu Wang}, {and} \bibinfo{person}{Xuanzhe Liu}.} \bibinfo{year}{2021}\natexlab{}.
\newblock \showarticletitle{{An Empirical Study on Deployment Faults of Deep Learning Based Mobile Applications}}. In \bibinfo{booktitle}{\emph{2021 IEEE/ACM 43rd International Conference on Software Engineering (ICSE)}}. IEEE, \bibinfo{pages}{674--685}.
\newblock


\bibitem[\protect\citeauthoryear{Chollet}{Chollet}{2015}]%
        {chollet2015keras}
\bibfield{author}{\bibinfo{person}{Fran\c{c}ois et~al. Chollet}.} \bibinfo{year}{2015}\natexlab{}.
\newblock \bibinfo{title}{Keras}.
\newblock \bibinfo{howpublished}{\url{https://keras.io}}.
\newblock


\bibitem[\protect\citeauthoryear{Deng, Dong, Socher, Li, Li, and Fei-Fei}{Deng et~al\mbox{.}}{2009}]%
        {deng2009imagenet}
\bibfield{author}{\bibinfo{person}{Jia Deng}, \bibinfo{person}{Wei Dong}, \bibinfo{person}{Richard Socher}, \bibinfo{person}{Li-Jia Li}, \bibinfo{person}{Kai Li}, {and} \bibinfo{person}{Li Fei-Fei}.} \bibinfo{year}{2009}\natexlab{}.
\newblock \showarticletitle{{Imagenet: A Large-Scale Hierarchical Image Database}}. In \bibinfo{booktitle}{\emph{2009 IEEE conference on computer vision and pattern recognition}}. \bibinfo{pages}{248--255}.
\newblock


\bibitem[\protect\citeauthoryear{Eniser, Gerasimou, and Sen}{Eniser et~al\mbox{.}}{2019}]%
        {deepfault}
\bibfield{author}{\bibinfo{person}{Hasan~Ferit Eniser}, \bibinfo{person}{Simos Gerasimou}, {and} \bibinfo{person}{Alper Sen}.} \bibinfo{year}{2019}\natexlab{}.
\newblock \showarticletitle{{DeepFault: Fault Localization for Deep Neural Networks}}. In \bibinfo{booktitle}{\emph{Fundamental Approaches to Software Engineering}}. \bibinfo{pages}{171--191}.
\newblock
\showISBNx{978-3-030-16722-6}


\bibitem[\protect\citeauthoryear{et~al.}{et~al.}{2014}]%
        {kruskallwallistestmethodology}
\bibfield{author}{\bibinfo{person}{Eva~Ostertagov{\'a} et al.}} \bibinfo{year}{2014}\natexlab{}.
\newblock \showarticletitle{Methodology and Application of the Kruskal-Wallis Test}.
\newblock \bibinfo{journal}{\emph{Applied Mechanics and Materials}}  \bibinfo{volume}{611} (\bibinfo{year}{2014}), \bibinfo{pages}{115 -- 120}.
\newblock
\urldef\tempurl%
\url{https://api.semanticscholar.org/CorpusID:119830984}
\showURL{%
\tempurl}


\bibitem[\protect\citeauthoryear{Gibson, Cano, Crowley, Storkey, and O'Boyle}{Gibson et~al\mbox{.}}{2025}]%
        {gibson_dlas_2025}
\bibfield{author}{\bibinfo{person}{Perry Gibson}, \bibinfo{person}{Jos\'e Cano}, \bibinfo{person}{Elliot.~J. Crowley}, \bibinfo{person}{Amos Storkey}, {and} \bibinfo{person}{Michael O'Boyle}.} \bibinfo{year}{2025}\natexlab{}.
\newblock \showarticletitle{{DLAS}: A Conceptual Model for Across-Stack Deep Learning Acceleration}.
\newblock \bibinfo{journal}{\emph{ACM Transactions on Architecture and Code Optimization (TACO)}} (\bibinfo{year}{2025}).
\newblock


\bibitem[\protect\citeauthoryear{He, Zhang, Ren, and Sun}{He et~al\mbox{.}}{2015}]%
        {resnet}
\bibfield{author}{\bibinfo{person}{Kaiming He}, \bibinfo{person}{Xiangyu Zhang}, \bibinfo{person}{Shaoqing Ren}, {and} \bibinfo{person}{Jian Sun}.} \bibinfo{year}{2015}\natexlab{}.
\newblock \showarticletitle{{Deep Residual Learning for Image Recognition}}.
\newblock \bibinfo{journal}{\emph{CoRR}}  \bibinfo{volume}{abs/1512.03385} (\bibinfo{year}{2015}).
\newblock
\showeprint[arXiv]{1512.03385}
\urldef\tempurl%
\url{http://arxiv.org/abs/1512.03385}
\showURL{%
\tempurl}


\bibitem[\protect\citeauthoryear{He, Zhang, Ren, and Sun}{He et~al\mbox{.}}{2016}]%
        {resnetv2}
\bibfield{author}{\bibinfo{person}{Kaiming He}, \bibinfo{person}{Xiangyu Zhang}, \bibinfo{person}{Shaoqing Ren}, {and} \bibinfo{person}{Jian Sun}.} \bibinfo{year}{2016}\natexlab{}.
\newblock \showarticletitle{Identity Mappings in Deep Residual Networks}.
\newblock \bibinfo{journal}{\emph{CoRR}}  \bibinfo{volume}{abs/1603.05027} (\bibinfo{year}{2016}).
\newblock
\showeprint[arXiv]{1603.05027}
\urldef\tempurl%
\url{http://arxiv.org/abs/1603.05027}
\showURL{%
\tempurl}


\bibitem[\protect\citeauthoryear{Islam, Pan, Nguyen, and Rajan}{Islam et~al\mbox{.}}{2020}]%
        {repairingDNNs}
\bibfield{author}{\bibinfo{person}{Md~Johirul Islam}, \bibinfo{person}{Rangeet Pan}, \bibinfo{person}{Giang Nguyen}, {and} \bibinfo{person}{Hridesh Rajan}.} \bibinfo{year}{2020}\natexlab{}.
\newblock \showarticletitle{Repairing Deep Neural Networks: Fix Patterns and Challenges}. In \bibinfo{booktitle}{\emph{Proceedings of the ACM/IEEE 42nd International Conference on Software Engineering}} \emph{(\bibinfo{series}{ICSE '20})}. \bibinfo{publisher}{Association for Computing Machinery}, \bibinfo{address}{New York, NY, USA}, \bibinfo{pages}{1135–1146}.
\newblock
\showISBNx{9781450371216}
\urldef\tempurl%
\url{https://doi.org/10.1145/3377811.3380378}
\showDOI{\tempurl}


\bibitem[\protect\citeauthoryear{Jajal, Jiang, Tewari, Kocinare, Woo, Sarraf, Lu, Thiruvathukal, and Davis}{Jajal et~al\mbox{.}}{2024}]%
        {InteroperabilityConverters}
\bibfield{author}{\bibinfo{person}{Purvish Jajal}, \bibinfo{person}{Wenxin Jiang}, \bibinfo{person}{Arav Tewari}, \bibinfo{person}{Erik Kocinare}, \bibinfo{person}{Joseph Woo}, \bibinfo{person}{Anusha Sarraf}, \bibinfo{person}{Yung-Hsiang Lu}, \bibinfo{person}{George~K. Thiruvathukal}, {and} \bibinfo{person}{James~C. Davis}.} \bibinfo{year}{2024}\natexlab{}.
\newblock \showarticletitle{Interoperability in Deep Learning: A User Survey and Failure Analysis of ONNX Model Converters}. In \bibinfo{booktitle}{\emph{Proceedings of the 33rd ACM SIGSOFT International Symposium on Software Testing and Analysis}} \emph{(\bibinfo{series}{ISSTA 2024})}. \bibinfo{publisher}{Association for Computing Machinery}, \bibinfo{address}{New York, NY, USA}, \bibinfo{pages}{1466–1478}.
\newblock
\showISBNx{9798400706127}
\urldef\tempurl%
\url{https://doi.org/10.1145/3650212.3680374}
\showDOI{\tempurl}


\bibitem[\protect\citeauthoryear{Liu, Chen, Zhang, Qin, Ji, Lin, and Yang}{Liu et~al\mbox{.}}{2020}]%
        {MMdnn}
\bibfield{author}{\bibinfo{person}{Yu Liu}, \bibinfo{person}{Cheng Chen}, \bibinfo{person}{Ru Zhang}, \bibinfo{person}{Tingting Qin}, \bibinfo{person}{Xiang Ji}, \bibinfo{person}{Haoxiang Lin}, {and} \bibinfo{person}{Mao Yang}.} \bibinfo{year}{2020}\natexlab{}.
\newblock \showarticletitle{Enhancing the Interoperability between Deep Learning Frameworks by Model Conversion}. In \bibinfo{booktitle}{\emph{Proceedings of the 28th ACM Joint Meeting on European Software Engineering Conference and Symposium on the Foundations of Software Engineering}} \emph{(\bibinfo{series}{ESEC/FSE 2020})}. \bibinfo{publisher}{Association for Computing Machinery}, \bibinfo{address}{New York, NY, USA}, \bibinfo{pages}{1320–1330}.
\newblock
\showISBNx{9781450370431}
\urldef\tempurl%
\url{https://doi.org/10.1145/3368089.3417051}
\showDOI{\tempurl}


\bibitem[\protect\citeauthoryear{Louloudakis, Gibson, Cano, and Rajan}{Louloudakis et~al\mbox{.}}{2022}]%
        {louloudakis2022assessing}
\bibfield{author}{\bibinfo{person}{Nikolaos Louloudakis}, \bibinfo{person}{Perry Gibson}, \bibinfo{person}{Jos{\'e} Cano}, {and} \bibinfo{person}{Ajitha Rajan}.} \bibinfo{year}{2022}\natexlab{}.
\newblock \showarticletitle{{Assessing Robustness of Image Recognition Models to Changes in the Computational Environment}}. In \bibinfo{booktitle}{\emph{NeurIPS ML Safety Workshop}}.
\newblock
\urldef\tempurl%
\url{https://openreview.net/forum?id=-7DjNGvdpx}
\showURL{%
\tempurl}


\bibitem[\protect\citeauthoryear{Louloudakis, Gibson, Cano, and Rajan}{Louloudakis et~al\mbox{.}}{2023a}]%
        {louloudakis2023deltann}
\bibfield{author}{\bibinfo{person}{Nikolaos Louloudakis}, \bibinfo{person}{Perry Gibson}, \bibinfo{person}{Jos{\'e} Cano}, {and} \bibinfo{person}{Ajitha Rajan}.} \bibinfo{year}{2023}\natexlab{a}.
\newblock \showarticletitle{{DeltaNN}: Assessing the Impact of Computational Environment Parameters on the Performance of Image Recognition Models}. In \bibinfo{booktitle}{\emph{39th IEEE International Conference on Software Maintenance and Evolution}}. \bibinfo{pages}{1--11}.
\newblock


\bibitem[\protect\citeauthoryear{Louloudakis, Gibson, Cano, and Rajan}{Louloudakis et~al\mbox{.}}{2023b}]%
        {louloudakis2023exploringeffectscomputationalparameter}
\bibfield{author}{\bibinfo{person}{Nikolaos Louloudakis}, \bibinfo{person}{Perry Gibson}, \bibinfo{person}{José Cano}, {and} \bibinfo{person}{Ajitha Rajan}.} \bibinfo{year}{2023}\natexlab{b}.
\newblock \bibinfo{title}{Exploring Effects of Computational Parameter Changes to Image Recognition Systems}.
\newblock
\newblock
\showeprint[arxiv]{cs.LG/2211.00471}
\urldef\tempurl%
\url{https://arxiv.org/abs/2211.00471}
\showURL{%
\tempurl}


\bibitem[\protect\citeauthoryear{Louloudakis, Gibson, Cano, and Rajan}{Louloudakis et~al\mbox{.}}{2024}]%
        {LouloudakisBuggyConversions}
\bibfield{author}{\bibinfo{person}{Nikolaos Louloudakis}, \bibinfo{person}{Perry Gibson}, \bibinfo{person}{Jos\'{e} Cano}, {and} \bibinfo{person}{Ajitha Rajan}.} \bibinfo{year}{2024}\natexlab{}.
\newblock \showarticletitle{Fault Localization for Buggy Deep Learning Framework Conversions in Image Recognition}. In \bibinfo{booktitle}{\emph{Proceedings of the 38th IEEE/ACM International Conference on Automated Software Engineering}} \emph{(\bibinfo{series}{ASE '23})}. \bibinfo{publisher}{IEEE Press}, \bibinfo{pages}{1795–1799}.
\newblock
\showISBNx{9798350329964}
\urldef\tempurl%
\url{https://doi.org/10.1109/ASE56229.2023.00147}
\showDOI{\tempurl}


\bibitem[\protect\citeauthoryear{{Mart\'{i}n~Abadi et al.}}{{Mart\'{i}n~Abadi et al.}}{2015}]%
        {tensorflow2015-whitepaper}
\bibfield{author}{\bibinfo{person}{{Mart\'{i}n~Abadi et al.}}} \bibinfo{year}{2015}\natexlab{}.
\newblock \bibinfo{title}{{{TensorFlow}: Large-Scale Machine Learning on Heterogeneous Systems}}.
\newblock
\newblock
\urldef\tempurl%
\url{https://www.tensorflow.org/}
\showURL{%
\tempurl}
\newblock
\shownote{{Software available from tensorflow.org}.}


\bibitem[\protect\citeauthoryear{Openja, Nikanjam, Yahmed, Khomh, and Jiang}{Openja et~al\mbox{.}}{2022a}]%
        {dlfconversionsstudy}
\bibfield{author}{\bibinfo{person}{Moses Openja}, \bibinfo{person}{Amin Nikanjam}, \bibinfo{person}{Ahmed~Haj Yahmed}, \bibinfo{person}{Foutse Khomh}, {and} \bibinfo{person}{Zhen Ming~Jack Jiang}.} \bibinfo{year}{2022}\natexlab{a}.
\newblock \showarticletitle{{An Empirical Study of Challenges in Converting Deep Learning Models}}. In \bibinfo{booktitle}{\emph{2022 IEEE International Conference on Software Maintenance and Evolution (ICSME)}}. \bibinfo{pages}{13--23}.
\newblock
\urldef\tempurl%
\url{https://doi.org/10.1109/ICSME55016.2022.00010}
\showDOI{\tempurl}


\bibitem[\protect\citeauthoryear{Openja, Nikanjam, Yahmed, Khomh, and Jiang}{Openja et~al\mbox{.}}{2022b}]%
        {openja2022empirical}
\bibfield{author}{\bibinfo{person}{Moses Openja}, \bibinfo{person}{Amin Nikanjam}, \bibinfo{person}{Ahmed~Haj Yahmed}, \bibinfo{person}{Foutse Khomh}, {and} \bibinfo{person}{Zhen Ming~Jack Jiang}.} \bibinfo{year}{2022}\natexlab{b}.
\newblock \showarticletitle{An empirical study of challenges in converting deep learning models}. In \bibinfo{booktitle}{\emph{2022 IEEE International Conference on Software Maintenance and Evolution (ICSME)}}. IEEE, \bibinfo{pages}{13--23}.
\newblock


\bibitem[\protect\citeauthoryear{Papineni, Roukos, Ward, and Zhu}{Papineni et~al\mbox{.}}{2002}]%
        {BLEU}
\bibfield{author}{\bibinfo{person}{Kishore Papineni}, \bibinfo{person}{Salim Roukos}, \bibinfo{person}{Todd Ward}, {and} \bibinfo{person}{Wei-Jing Zhu}.} \bibinfo{year}{2002}\natexlab{}.
\newblock \showarticletitle{BLEU: a method for automatic evaluation of machine translation}. In \bibinfo{booktitle}{\emph{Proceedings of the 40th Annual Meeting on Association for Computational Linguistics}} \emph{(\bibinfo{series}{ACL '02})}. \bibinfo{publisher}{Association for Computational Linguistics}, \bibinfo{address}{USA}, \bibinfo{pages}{311–318}.
\newblock
\urldef\tempurl%
\url{https://doi.org/10.3115/1073083.1073135}
\showDOI{\tempurl}


\bibitem[\protect\citeauthoryear{Paszke, Gross, Massa, Lerer, Bradbury, Chanan, Killeen, Lin, Gimelshein, Antiga, Desmaison, K{\"{o}}pf, Yang, DeVito, Raison, Tejani, Chilamkurthy, Steiner, Fang, Bai, and Chintala}{Paszke et~al\mbox{.}}{2019}]%
        {pytorch}
\bibfield{author}{\bibinfo{person}{Adam Paszke}, \bibinfo{person}{Sam Gross}, \bibinfo{person}{Francisco Massa}, \bibinfo{person}{Adam Lerer}, \bibinfo{person}{James Bradbury}, \bibinfo{person}{Gregory Chanan}, \bibinfo{person}{Trevor Killeen}, \bibinfo{person}{Zeming Lin}, \bibinfo{person}{Natalia Gimelshein}, \bibinfo{person}{Luca Antiga}, \bibinfo{person}{Alban Desmaison}, \bibinfo{person}{Andreas K{\"{o}}pf}, \bibinfo{person}{Edward~Z. Yang}, \bibinfo{person}{Zach DeVito}, \bibinfo{person}{Martin Raison}, \bibinfo{person}{Alykhan Tejani}, \bibinfo{person}{Sasank Chilamkurthy}, \bibinfo{person}{Benoit Steiner}, \bibinfo{person}{Lu Fang}, \bibinfo{person}{Junjie Bai}, {and} \bibinfo{person}{Soumith Chintala}.} \bibinfo{year}{2019}\natexlab{}.
\newblock \showarticletitle{{PyTorch: An Imperative Style, High-Performance Deep Learning Library}}.
\newblock \bibinfo{journal}{\emph{CoRR}}  \bibinfo{volume}{abs/1912.01703} (\bibinfo{year}{2019}).
\newblock
\showeprint[arXiv]{1912.01703}
\urldef\tempurl%
\url{http://arxiv.org/abs/1912.01703}
\showURL{%
\tempurl}


\bibitem[\protect\citeauthoryear{Pham, Lutellier, Qi, and Tan}{Pham et~al\mbox{.}}{2019}]%
        {pham2019cradle}
\bibfield{author}{\bibinfo{person}{Hung~Viet Pham}, \bibinfo{person}{Thibaud Lutellier}, \bibinfo{person}{Weizhen Qi}, {and} \bibinfo{person}{Lin Tan}.} \bibinfo{year}{2019}\natexlab{}.
\newblock \showarticletitle{{CRADLE: Cross-Backend Validation to Detect and Localize Bugs in Deep Learning Libraries}}. In \bibinfo{booktitle}{\emph{2019 IEEE/ACM 41st International Conference on Software Engineering (ICSE)}}. \bibinfo{pages}{1027--1038}.
\newblock
\urldef\tempurl%
\url{https://doi.org/10.1109/ICSE.2019.00107}
\showDOI{\tempurl}


\bibitem[\protect\citeauthoryear{Recht, Roelofs, Schmidt, and Shankar}{Recht et~al\mbox{.}}{2019}]%
        {imagenetv2}
\bibfield{author}{\bibinfo{person}{Benjamin Recht}, \bibinfo{person}{Rebecca Roelofs}, \bibinfo{person}{Ludwig Schmidt}, {and} \bibinfo{person}{Vaishaal Shankar}.} \bibinfo{year}{2019}\natexlab{}.
\newblock \bibinfo{title}{Do ImageNet Classifiers Generalize to ImageNet?}
\newblock
\newblock
\showeprint[arxiv]{cs.CV/1902.10811}


\bibitem[\protect\citeauthoryear{Russakovsky, Deng, Su, Krause, Satheesh, Ma, Huang, Karpathy, Khosla, Bernstein, Berg, and Fei-Fei}{Russakovsky et~al\mbox{.}}{2015}]%
        {ILSVRC17}
\bibfield{author}{\bibinfo{person}{Olga Russakovsky}, \bibinfo{person}{Jia Deng}, \bibinfo{person}{Hao Su}, \bibinfo{person}{Jonathan Krause}, \bibinfo{person}{Sanjeev Satheesh}, \bibinfo{person}{Sean Ma}, \bibinfo{person}{Zhiheng Huang}, \bibinfo{person}{Andrej Karpathy}, \bibinfo{person}{Aditya Khosla}, \bibinfo{person}{Michael Bernstein}, \bibinfo{person}{Alexander~C. Berg}, {and} \bibinfo{person}{Li Fei-Fei}.} \bibinfo{year}{2015}\natexlab{}.
\newblock \showarticletitle{{ImageNet Large Scale Visual Recognition Challenge}}.
\newblock \bibinfo{journal}{\emph{International Journal of Computer Vision (IJCV)}} \bibinfo{volume}{115}, \bibinfo{number}{3} (\bibinfo{year}{2015}), \bibinfo{pages}{211--252}.
\newblock
\urldef\tempurl%
\url{https://doi.org/10.1007/s11263-015-0816-y}
\showDOI{\tempurl}


\bibitem[\protect\citeauthoryear{Sandler, Howard, Zhu, Zhmoginov, and Chen}{Sandler et~al\mbox{.}}{2018}]%
        {mobilenetv2}
\bibfield{author}{\bibinfo{person}{Mark Sandler}, \bibinfo{person}{Andrew~G. Howard}, \bibinfo{person}{Menglong Zhu}, \bibinfo{person}{Andrey Zhmoginov}, {and} \bibinfo{person}{Liang{-}Chieh Chen}.} \bibinfo{year}{2018}\natexlab{}.
\newblock \showarticletitle{{Inverted Residuals and Linear Bottlenecks: Mobile Networks for Classification, Detection and Segmentation}}.
\newblock \bibinfo{journal}{\emph{CoRR}}  \bibinfo{volume}{abs/1801.04381} (\bibinfo{year}{2018}).
\newblock
\showeprint[arXiv]{1801.04381}
\urldef\tempurl%
\url{http://arxiv.org/abs/1801.04381}
\showURL{%
\tempurl}


\bibitem[\protect\citeauthoryear{Szegedy, Vanhoucke, Ioffe, Shlens, and Wojna}{Szegedy et~al\mbox{.}}{2015}]%
        {inceptionv3}
\bibfield{author}{\bibinfo{person}{Christian Szegedy}, \bibinfo{person}{Vincent Vanhoucke}, \bibinfo{person}{Sergey Ioffe}, \bibinfo{person}{Jonathon Shlens}, {and} \bibinfo{person}{Zbigniew Wojna}.} \bibinfo{year}{2015}\natexlab{}.
\newblock \showarticletitle{{Rethinking the Inception Architecture for Computer Vision}}.
\newblock \bibinfo{journal}{\emph{CoRR}}  \bibinfo{volume}{abs/1512.00567} (\bibinfo{year}{2015}).
\newblock
\showeprint[arXiv]{1512.00567}
\urldef\tempurl%
\url{http://arxiv.org/abs/1512.00567}
\showURL{%
\tempurl}


\bibitem[\protect\citeauthoryear{Wang, Yan, Chen, Liu, and Zhang}{Wang et~al\mbox{.}}{2020}]%
        {wang2020lemon}
\bibfield{author}{\bibinfo{person}{Zan Wang}, \bibinfo{person}{Ming Yan}, \bibinfo{person}{Junjie Chen}, \bibinfo{person}{Shuang Liu}, {and} \bibinfo{person}{Dongdi Zhang}.} \bibinfo{year}{2020}\natexlab{}.
\newblock \showarticletitle{{Deep Learning Library Testing via Effective Model Generation}}. In \bibinfo{booktitle}{\emph{Proceedings of the 28th ACM Joint Meeting on European Software Engineering Conference and Symposium on the Foundations of Software Engineering}}. \bibinfo{pages}{788–799}.
\newblock
\showISBNx{9781450370431}


\bibitem[\protect\citeauthoryear{Wardat, Le, and Rajan}{Wardat et~al\mbox{.}}{2021}]%
        {wardat2021deeplocalize}
\bibfield{author}{\bibinfo{person}{Mohammad Wardat}, \bibinfo{person}{Wei Le}, {and} \bibinfo{person}{Hridesh Rajan}.} \bibinfo{year}{2021}\natexlab{}.
\newblock \showarticletitle{{DeepLocalize: Fault Localization for Deep Neural Networks}}. In \bibinfo{booktitle}{\emph{Proceedings of the 43rd International Conference on Software Engineering}}. \bibinfo{pages}{251–262}.
\newblock
\showISBNx{9781450390859}


\end{thebibliography}

\end{document}